\documentclass[a4paper,twocolumn,10pt,accepted=2024-05-14]{quantumarticle}
\pdfoutput=1
\usepackage[numbers,sort&compress]{natbib}
\usepackage{placeins}
\usepackage[margin=20mm]{geometry}
\usepackage{xcolor}
\usepackage{amssymb,amsmath,amstext}
\usepackage{amsfonts}
\usepackage{braket}
\usepackage{pgfplots}
\usepackage{mathrsfs}
\usepackage{physics}
\usepackage[caption = false]{subfig}
\usepackage{bbold}
\usepackage{bbm}
\usepackage[normalem]{ulem}
\usepackage{latexsym}
\usepackage{umoline}
\usepackage{dsfont}
\usepackage{comment}
\usepackage{float}
\usepackage{hhline}
\usepackage{graphicx}
\usepackage{lipsum}
\usepackage{float}
\usepackage{hyperref}
\usepackage{mathrsfs}
\usepackage{footnote}

\graphicspath{{./figures/}}

\DeclareMathOperator{\Cc}{\mathit{C}}

\pgfplotsset{compat=1.17}

\begin{document}

\title{The Bethe Ansatz as a Quantum Circuit}

\author{Roberto Ruiz}
\affiliation{Instituto de F\'{\i}sica Te\'{o}rica, UAM/CSIC, Universidad Aut\'{o}noma de Madrid, Madrid, Spain}
\author{Alejandro Sopena}
\affiliation{Instituto de F\'{\i}sica Te\'{o}rica, UAM/CSIC, Universidad Aut\'{o}noma de Madrid, Madrid, Spain}
\author{Max Hunter Gordon}
\affiliation{Instituto de F\'{\i}sica Te\'{o}rica, UAM/CSIC, Universidad Aut\'{o}noma de Madrid, Madrid, Spain}
\affiliation{Normal Computing Corporation, New York, New York, USA}
\author{Germ\'{a}n Sierra}
\affiliation{Instituto de F\'{\i}sica Te\'{o}rica, UAM/CSIC, Universidad Aut\'{o}noma de Madrid, Madrid, Spain}
\author{Esperanza L\'{o}pez}
\affiliation{Instituto de F\'{\i}sica Te\'{o}rica, UAM/CSIC, Universidad Aut\'{o}noma de Madrid, Madrid, Spain}

\begin{abstract}
The Bethe ansatz represents an analytical method enabling the exact solution of numerous models in condensed matter physics and statistical mechanics. When a global symmetry is present, the trial wavefunctions of the Bethe ansatz consist of plane wave superpositions. Previously, it has been shown that the Bethe ansatz can be recast as a deterministic quantum circuit. 
An analytical derivation of the quantum gates that form the circuit was lacking however.
Here we present a comprehensive study of the transformation that brings the Bethe ansatz into a quantum circuit, which leads us to determine the analytical expression of the circuit gates. 
As a crucial step of the derivation, we present a simple set of diagrammatic rules that define a novel Matrix Product State network building Bethe wavefunctions. 
Remarkably, this provides a new perspective on the equivalence between the coordinate and algebraic versions of the Bethe ansatz.

\end{abstract}
\maketitle

\section{Introduction}

In 1931 Hans Bethe introduced an ansatz for the eigenstates of the antiferromagnetic Heisenberg Hamiltonian in a closed  spin-$1/2$ chain~\cite{Bethe31}. 
This ansatz involves the summation of permutations of plane waves, with their quasi-momenta coupled through
transcendental equations known as Bethe equations. 
This groundbreaking study  marked the inception of exactly 
solvable models in 
condensed matter physics and statistical mechanics 
~\cite{Baxter82,pink,Mussardo20}.
Subsequently, the Bethe ansatz was expanded to encompass 
the XXZ model~\cite{Yang66}, 
the one-dimensional Bose gas with delta function interactions~\cite{Lieb63}, 
the Hubbard model~\cite{Lieb68}, 
and a plethora of other systems~\cite{Korepin1993kvr, Sutherland04, Gaudin14, Gohmann23}.

At the core of the exact solutions for these models lies their integrability, which denotes the presence 
of a maximal number of conserved quantities~\cite{Caux10}. 
The latter naturally emerge within the framework of the Algebraic Bethe Ansatz (ABA), 
pioneered by the Faddeev's school in the 70s--80s~\cite{faddeev1996algebraic}. 
This approach unveils the algebraic underpinnings of the Bethe ansatz, 
which are based on an $R$ matrix that fulfils the Yang-Baxter equation. 

In recent years the study of quantum many body systems has been strongly influenced by quantum information theory and quantum computing. 
In turn, this has led to a more profound understanding of the structure of many body wavefunctions, in particular with regards to entanglement structure. 
It was discovered that tensor networks provide the natural language with which to express entanglement structure in many body wavefunctions. 
A one-dimensional tensor network is a Matrix Product State (MPS), for a summary see~\cite{Cirac09,Orus13,Cirac21}. 
In view of these developments, it was natural to revisit the Bethe ansatz and explore how it could be reframed in the language of tensor networks. 
This was undertaken in~\cite{Alcaraz03,Alcaraz03ii,Alcaraz06}, 
where a huge variety of exact integrable quantum chains were expressed in a unified way using matrix product ansatze, that are essentially equivalent to the ABA~\cite{Katsura10, Murg_2012}.

In parallel, the quantum computing community pursued the construction of efficient quantum circuits to prepare many body wavefunctions on quantum computers, 
taking advantage of the exact solutions derived for models such as the XY and Ising models. This was achieved either by the mapping of these models to free fermions~\cite{Verstraete_2009,CerveraLierta2018exactisingmodel} or by the elaboration on spin variables directly~\cite{Kivlichan18}. 
However, the challenge of determining how to construct quantum circuits for models that involve interactions remains. In this work, we directly address this problem. We do so by considering a highly pertinent question: 
could the Bethe ansatz, which describes interacting models, be adapted to the newly introduced quantum computers?~\cite{Nepomechie21}. 
Among the reasons behind this inquiry is the potential to compute traditionally inaccessible quantities via measurement, 
such as correlation functions with arbitrary ranges and higher orders~\cite{Balazs2017}. This might lead to a quantum advantage within this particular framework, which could be implemented in the currently available quantum computers. It is worth noting that low-energy states of integrable gapped spin chains can be prepared with MPS efficiently~\cite{Cirac09,Orus13,Cirac21,Malz23}. However, both highly excited states of gapped spin chains and general states of ungapped integrable spin chains are out of reach for classical simulatability.

Progress in this direction was made in~\cite{VanDyke21,VanDyke22}, where an algorithm to prepare eigenstates of the XXZ model was introduced. The algorithm complexity was polynomial in both the number of $T$ gates and circuit depth. However, its probabilistic nature resulted in the success rate to diminish exponentially~\cite{Li22}. Its applicability was moreover confined to real valued solutions of the Bethe equations, which presented a significant limitation.

A different pathway for creating a quantum circuit linked to the Bethe ansatz was introduced in~\cite{Sopena22}, by using its algebraic representation. 
In contrast to the previously discussed approach rooted in the original plane wave ansatz, this alternative method leverages viewing the ABA as an MPS. 
Consequently, it could be transformed into a quantum circuit 
using conventional techniques from the field of tensor networks. 
The resulting circuit, termed Algebraic Bethe Circuit (ABC), 
offer the advantages of being deterministic and applicable to any solution, 
whether real or complex, of the Bethe equations. 

ABCs are formulated in terms of multiqubit quantum gates. The algorithm complexity therefore translates into the efficiency of their decomposition in terms of one-qubit and two-qubit unitaries. While an efficient decomposition was proposed for the noninteracting XX chain, no conclusive answer was found for the XXZ model. 
The difficulty inherent to this problem was exacerbated by 
the lack of analytical expressions in a closed form for the gates comprising the circuit. 
Indeed, closed expressions were only obtained for scenarios involving one or two magnons, while more complicated cases were addressed numerically.

In this paper we comprehensively study the transformation leading to the ABC and arrive at an analytical expression for the quantum gates that form the ABC. Instead of relying on the MPS interpretation of the ABA, we use a simpler MPS structure underlying the Bethe ansatz. This structure turns out to be naturally linked to Bethe plane wave ansatz, also known as Coordinate Bethe Ansatz (CBA). Remarkably, the CBA related MPS can be encoded in a set of diagrammatic rules, which we exploit to present our findings. The shift from the ABA to the CBA ultimately enables us to provide complete analytical expressions for the unitary matrices that comprise the quantum circuit. These results are summarized in the following figure 
\begin{equation}
\vcenter{\hbox{\includegraphics[width=.9\columnwidth]{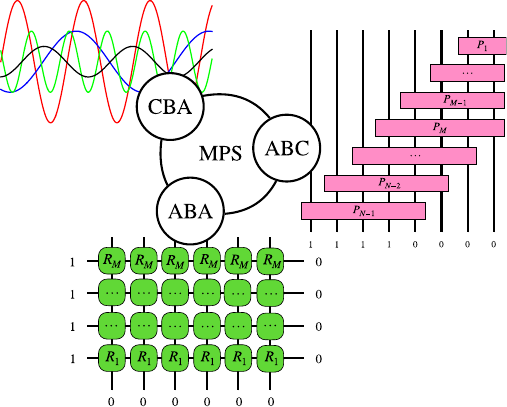}}}  
\label{resume}
\end{equation} 
The equivalence between the ABA and the CBA was proven soon after the systematic analysis of quantum integrability began. This equivalence, which holds independently of the Bethe equations, is highly nontrivial~\cite{Sklyanin82}. 
Interestingly, the MPS reformulation of the coordinate and algebraic versions of the Bethe ansatz will lead us to novel insights into their equivalence, from which we recover the integrability constraints contained in the Yang-Baxter equation.

The paper is organized as follows: Section~\ref{sec:ABC} provides a review of the Bethe wavefunction for the XXZ model, its representation through the ABA, and introduces the quantum circuit structure outlined in~\cite{Sopena22}.
Section~\ref{sec:Symmetry} presents the Hilbert space structure  based on the $U(1)$ symmetry of the model, 
highlighting the resulting gate decomposition within the quantum circuit. 
Section~\ref{sec:Ansatz} introduces the proposed analytic form of the general ABC gates,
drawing inspiration from the simple one-magnon case. 
Section~\ref{sec:Lambda} introduces a set of diagrammatic rules defining an MPS that builds Bethe wavefunctions. Section~\ref{sec:A_B} implements the transformation of this MPS into canonical form ~\cite{Cirac09,Orus13,Cirac21}, such that valid quantum gates can be distilled. Section~\ref{sec:short_gates} shows how to obtain a deterministic quantum circuit. Section~\ref{sec:Unitarity} offers a proof of the unitarity of the proposed circuit. Section~\ref{sec:ABACBA} proposes an alternative understanding of the equivalence between the ABA and the CBA. 
Section~\ref{sec:free_fermion} specializes the construction to the XX model, and presents an efficient decomposition of the associated ABC gates into two-qubit unitaries. Section~\ref{sec:Conclusions}  concludes by summarising key findings and contributions. Several appendices explain technical issues.

\section{\label{sec:ABC}Algebraic Bethe Circuits}

In this section we introduce the trial wavefunctions that are at the basis of the Bethe ansatz, and review the construction of their quantum circuit representation proposed in~\cite{Sopena22}.

\subsection{Bethe wavefunctions}

The trial wavefunctions of the Bethe ansatz are linear superpositions of spin waves, also called magnons. 
Their construction relies upon the presence of a $U(1)$ symmetry whose conserved quantity is the projection of
the total spin along a chosen direction, which customarily
is taken to be the $z$ axis. The $U(1)$ symmetry implies the conservation of the total number of magnons. Due to integrability, the interaction among magnons factorizes in two-body scattering events that preserve the value of their individual momenta.

Throughout the paper we focus on a spin-$1/2$ chain. Together with the requirements of integrability, the presence of $U(1)$ symmetry and nearest neighbours interaction, this leads to the XXZ Heisenberg model described by the Hamiltonian
\begin{equation}
H = \sum_{j=1}^N \,(\sigma^x_j \sigma^x_{j+1} + \sigma^y_j \sigma^y_{j+1} + \Delta \, \sigma^z_j \sigma^z_{j+1}) \ ,
\label{ham}
\end{equation}
where $\sigma^\alpha_j \; (\alpha=x,y,z)$ are the Pauli matrices acting at the site $j=1, \dots, N$, and 
$\sigma_{N+1}^\alpha = \sigma^\alpha_1$ for periodic boundary conditions. 
One spin-$1/2$ site and one qubit share the same Hilbert space, namely, $\mathcal{H}=\mathbb{C}^2$. 
We will then identify the spin basis with the computational basis of a qubit by 
\begin{equation}
\ket{\uparrow}\equiv\ket{0} \ , \quad \ket{\downarrow}\equiv\ket{1} \ .
\end{equation}
The $U(1)$ symmetry allows us to choose $| 0_N\rangle=|0\rangle^{\otimes N}$ as a reference state, where $N$ is the total number of sites of the chain. 
Magnons are spin waves over this reference state, such that the number of magnons counts the amount of sites in the state $\ket{1}$.

Imposing periodic boundary conditions on the trial wavefunctions leads to the Bethe equations for the magnon momenta. In this paper, however, we will only be concerned with the preparation of the trial wavefunctions on a quantum computer. We refer to them as Bethe {\it wavefunctions}, as opposed to Bethe {\it eigenstates} which are the result of applying the Bethe equations. Therefore, the boundary conditions should be addressed separately.

As an example let us consider the Bethe wavefunction describing two magnons supported by $N$ qubits.
It reads 
\begin{equation}
\begin{split}
|\Psi^{(2)}_N \rangle  = & \!\! \underset{n_1<n_2}{\sum_{ n_{1,2}=1}^N} \!\!\Big( s_{21} \,  x_1^{n_1-1} x_2^{n_2-1} \\[-6mm]
& \hspace{1cm} -  s_{12} \, x_1^{n_2-1} x_2^{n_1-1} \Big)| n_1 n_2 \rangle \ ,
\end{split}
\label{B2wf}
\end{equation}
where we defined the variables $x_a=e^{i p_a}$, with $p_a$ the magnon momenta, and $\ket{n_1 n_2}$ denotes the state of the computational basis of $N$ qubits with $\ket{1}$ at positions~{$n_1$} and $n_2$.  The only input of the construction is the scattering amplitude 
\begin{equation}
\label{scatampl}
s_{12}=s(p_1,p_2) \ , \hspace{1cm} s_{21}=s(p_2,p_1) \ ,
\end{equation}
which encodes the scattering $S$ matrix between two magnons 
\begin{equation}
S(p_1,p_2)=-\frac{s_{12}}{s_{21}} \ .
\label{phaseshift}
\end{equation}
In the XXZ model the function $s_{12}$ is given by 
\begin{equation}
s_{12}=1+x_1 x_2-2\Delta \, x_2 
\ ,  
\label{sfunction}
\end{equation}
where $\Delta$ is the coupling constant appearing in  the Heisenberg Hamiltonian~\eqref{ham}.
If the scattering amplitude is symmetric, the $S$ matrix reduces to a sign flip, in which case there exists a Jordan-Wigner transformation that maps the system to free fermions~\cite{Lieb61}. This result is consistent with the antisymmetry of the wavefunction under the exchange of momenta $p_1$ and $p_2$. 

For $M$ magnons, the Bethe wavefunction is 
\begin{equation}
\label{Bwf}
\begin{split}
|\Psi_N^{(M)} \rangle = \underset{n_m <n_{m\!+\!1}}{\sum_{ n_m=1 }^N}   \sum_{a_m=1}^M  \epsilon_{a_1 \ldots a_M} \underset{p>q}{\prod_{p,q=1  }^M} \! \!s_{a_{p} a_q}    \\
\times \; x_{a_1}^{n_{1}-1} \!\!\! \ldots  \, x_{a_M}^{n_{M}-1}  \; | n_1 n_2\ldots n_M \rangle  \ ,
\end{split}
\end{equation}
with $n_m$ signalling the position of the qubits at $\ket{1}$. The choice of ansatz~\eqref{Bwf}  is suitable for  the consideration of periodic boundary conditions.
By the assumption
of two-body factorization, the coefficients are proportional to the product of phase shifts~\eqref{scatampl} corresponding to the permutation that brings $1,\ldots,M$ into $a_1,\ldots,a_M$. 
Note that~\eqref{Bwf} is antisymmetric under permutations of any pair of momenta, and thus vanishes whenever two of them coincide. 
It should be stressed that the state~\eqref{Bwf} is not normalized. 
We shall return to this fact in due course. 
The previous construction of the wavefunction is known as Coordinate Bethe Ansatz, or CBA for short. 

\subsection{From ABA to a quantum circuit}

As explained in the introduction, there exists an algebraic approach to compute Bethe wavefunctions, termed Algebraic Bethe Ansatz or simply ABA, which exploits the quantum integrability of the system explicitly. The wavefunction in the ABA is built upon the action of magnon creation operators $B(p)$ on
the reference state
\begin{equation}
| \Psi_N^{(M)} \rangle =  \rho \,
B(p_1) \dots B(p_M) |0_N \rangle \ .
\label{BABA}
\end{equation}
The proportionality factor $\rho$, which in general depends on the momenta $p_a$, makes it clear that the norm of the states produced by the CBA and the ABA need not coincide.
The operators $B(p_a)$ admit a tensor network representation in terms of the 
$R$ matrix. The $R$ matrix defines a map on the tensor product of the Hilbert space of a site,  ${\cal H}_{\mathrm{phys}}$, and an auxiliary Hilbert space, ${\cal H}_{\mathrm{aux}}$. The role of the latter is to input the required number of $\ket{1}$ states into the network. Graphically, the ABA tensor network is given by
\begin{equation}
\vcenter{\hbox{\includegraphics[width=.8\columnwidth]{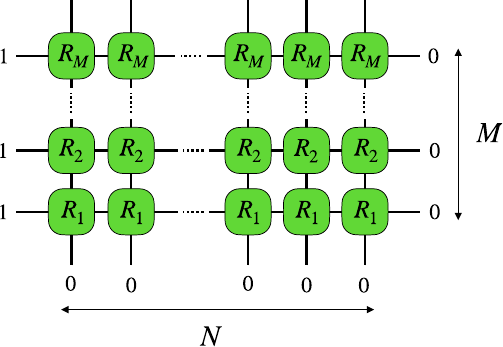}}}
\label{intro1}
\end{equation} 
Rounded boxes denote that the $R$ matrices are nonunitary. 
Each $R$ matrix depends on one magnon momentum,~\mbox{$R_a=R(p_a)$}, and each row of $R$ matrices, all of them depending on the same momentum, defines the operator $B(p_a)$.
This network can be recast in the circuitlike form
\vspace{-4mm}
\begin{equation}
\vcenter{\hbox{\includegraphics[width=.7\columnwidth]{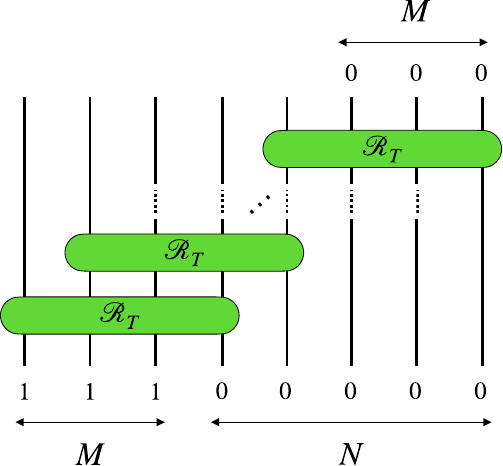}}} 
\label{ABAcircuit}
\end{equation} 
where the tensors $\mathscr{R}_T$ follow from the contraction of the~$R$ matrices in a column of~\eqref{intro1}. This reinterpretation requires $N$ physical   
plus $M$ ancillary qubits. The output state of every ancillary qubit must be set to $|0\rangle$ to recover the correct state~\eqref{BABA} in the physical qubits. 
In spite of the suggestive form, neither $R$ nor the $\mathscr{R}_T$ are unitary, and thus do not define quantum gates.

In~\cite{Sopena22}, a procedure was presented to distil unitary gates out of $\mathscr{R}_T$, and at the same time eliminate the ancillary qubits.
The procedure eventually supplies the following deterministic quantum circuit 
\begin{equation}
\vcenter{\hbox{\includegraphics[width=.66\columnwidth]{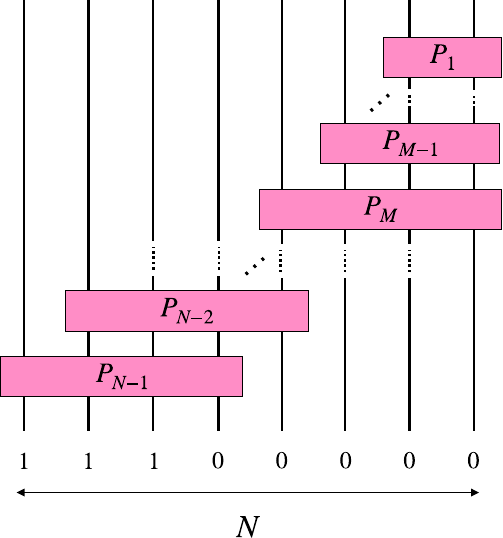}}}  
\label{ABCgraph}
\end{equation} 
which was named Algebraic Bethe Circuit or ABC.
Sharp boxes denote the gates are unitary. The gates $P_{k}$ are the solution to certain recursion relations involving $\mathscr{R}_T$. 
These recursion relations were solved numerically in~\cite{Sopena22}, although analytical expressions were obtained for wavefunctions describing one and two magnons.
In this paper we follow a different approach. Instead of focusing on the recursion relations, we will propose an informed ansatz for the ABC gates. A partial proof and strong numerical arguments will be then given in favour of its validity.
In this way, we derive the complete solution of the quantum circuit preparing 
the wavefunction~\eqref{Bwf}
for an arbitrary number of magnons and sites.  Although integrability motivates~\eqref{Bwf} and determines the scattering amplitude to have the form~\eqref{sfunction}, we note that the construction presented in the next sections applies for an arbitrary scattering amplitude  $s_{12}$.

\section{\label{sec:Symmetry}Symmetry sectors}

We introduce here our notation, which relies on the assumed $U(1)$ symmetry of the integrable chain.
We first present an efficient way to index the computational basis states belonging to different symmetry sectors. We then exploit both the symmetry and the staircase structure of the ABC to break down the $P_k$ gates into a collection of restricted maps.

\subsection{\label{amap}Hilbert space}

The Hilbert space of $k$ qubits, ${\cal H}_{k} =  (\mathbb{C}^2)^{\otimes{k}}$, is spanned by the computational basis
\begin{equation}
{\cal H}_k = {\rm Span} \;  \Big\{  |i_1\dots i_k \rangle \;  :  \;   i_j= 0,1 \Big\} \ .
\label{H01}
\end{equation}
We stress that we order qubits from left to right. 
The Hilbert space $\mathcal{H}_k$ decomposes into symmetry sectors with definite total spin along the $z$ axis, which, by the assumption of $U(1)$ symmetry, will be preserved by the quantum gates. The decomposition is~\mbox{${\cal H}_k  = \overset{k}{\underset{r=0}{\oplus}}{\cal H}_{r,k}$}, 
where ${\cal H}_{r,k}$ is the subspace with~$r$ qubits at $|1\rangle$,
\begin{equation}
{\cal H}_{r,k} = {\rm Span} \;  \Big\{  |i_1 \dots i_k \rangle \;  :  \;   \sum_{j=1}^k i_j = r  \Big\} \, . 
\label{hrn}
\end{equation}
The restriction of the computational basis to the $r$ symmetry sector can be equivalently described with the notation $\ket{n_1\dots n_r}$ of Section~\ref{sec:ABC}, which singles out the location of qubits at $\ket{1}$
\begin{equation}
\vcenter{\hbox{\includegraphics[width=.59\columnwidth]{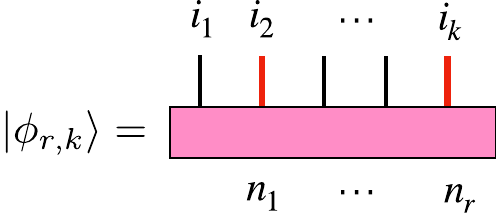}}}  
\label{gates0}
\end{equation} 
When we need to represent graphically the state of the qubits, we 
shall use a red line for $\ket{1}$ and a black line for~$\ket{0}$. 
The dimension of ${\cal H}_{r,k}$ is the binomial coefficient
\begin{equation}
 d_{r,k} =\binom{k}{r}= \frac{k!}{r!(k-r)!} \ , 
 \end{equation}
whose sum, 
using the binomial expansion, gives the dimension of ${\cal H}_k$.  

To describe the symmetry structure of the quantum gates, it would be convenient to 
label the basis states of each sector with an integer $a=1,\dots,d_{r,k}$. This is easily achieved as follows. Let the tuple~$(i_1\dots i_k)_r$ describe a computational basis state belonging to $ {\cal H}_{ r,k} $.
We can use its
binary digits to define an integer
\begin{equation}
   (i_1 \dots i_k)_r\; \rightarrow \, \chi= \sum_{j=1}^{k}\,  i_j \; 2^{j-1} \ .
\end{equation}
We sort every possible $\chi$ in an ordered list and assign to each $\chi$ the number $a$ that indexes its position in the list.   
We illustrate the map with the example $k=4$ and~{$r=2$} in the table below.

\begin{equation}
\begin{array}{cccc|c|c|cc}
i_1 & i_2 & i_3 & i_4 & \;\;\chi \;\;& \; a \; & n_1 & n_2 \\
\hline
1 & 1 & 0 & 0 & 3 & 1 & 1 & 2  \\
1 & 0 & 1 & 0 & 5 & 2 & 1 & 3  \\
1 & 0 & 0 & 1 & 9 & 4 & 1 & 4  \\
0 & 1 & 1 & 0 & 6 & 3 & 2 & 3  \\
0 & 1 & 0 & 1 & 10 & 5 & 2 & 4  \\
0 & 0 & 1 & 1 & 12 & 6 & 3 & 4  \\
\hline 
\end{array}
\label{eq:Table1}
\end{equation}

\subsection{\label{sec:Gate_structure}Gate structure}

The quantum circuit~\eqref{ABCgraph}
creates a Bethe wavefunction supported by $N$ qubits. It contains $N-1$ gates $P_k$, which can be classified in two types. 
When $k$ is smaller than the number of magnons, $M$, the gates $P_k$ act on a number of qubits that grows as $k+1$. We will refer to them as short gates.  They are crucial to obtain a deterministic quantum circuit. On the contrary, the  gates with $k \geq M$ act on a fixed number of $M+1$ qubits. We will call them  long gates.

First we describe the symmetry structure of the long gates.
Their rightmost input qubit is by construction always at $|0\rangle$, as can be seen in~\eqref{ABCgraph}. 
Hence, for our purposes, only the map
\begin{equation}
P_k |0 \rangle : {\cal H}_M\longrightarrow {\cal H}_{M+1} \ 
\label{Pdec}
\end{equation}
is relevant. Due to the underlying symmetry, this map decomposes as  
$P_k|0 \rangle = \oplus_{r=0}^M P^{(r)}_k$, where $P^{(r)}_k$
is the restriction of~\eqref{Pdec} to input and output configurations with~$r$ qubits in $\ket{1}$. In components, it reads
\vspace{1mm}
\begin{equation}
P_k |0 \rangle = 
\left( \begin{array}{cccccc}
1  & 0  & \cdots & 0 \\
0  & P^{(1)}_k & \cdots  & 0 \\
\vdots & \vdots & & \vdots \\[1mm]
0& 0  & \cdots & P^{(M)}_k  \\[1mm]
0& 0  & \cdots & 0 \\
\end{array}
\right) \, .
\label{bigP}
\end{equation}
\vspace{1mm}

\noindent
Without loss of generality, we have chosen $P_k^{(0)}=1$. The last row vanishes because there is no input state
in the symmetry sector $r=M+1$.

The small ABC gates are $k+1$ qubit unitaries with no restriction on the input configurations.
Therefore we need to determine the complete map
\begin{equation}
P_k : {\cal H}_{k+1} \longrightarrow {\cal H}_{k+1} \  .
\label{Pdec2}
\end{equation}
Breaking down into components restricted to different symmetry sectors, we have $P_k= \oplus_{r=0}^{k+1} P^{(r)}_k$, or equivalently 
\begin{equation}
    P_k = 
    \begin{pmatrix}
        1 & 0 &  \cdots & 0 &0 \\[1mm]
        0 & P_k^{(1)}  &\cdots & 0&0\\
        \vdots & \vdots  & & \vdots & \vdots  \\
        0 & 0 & \cdots & P_k^{(k)} & 0 \\[1mm]
        0 & 0 & \cdots & 0 & 1
    \end{pmatrix} \ ,
    \label{Pnfirst}
\end{equation}

\vspace{2mm}
\noindent where we have set $P_k^{(0)}=P_k^{(k+1)}=1$.

Up to now we have been exploiting the $U(1)$ symmetry of the integrable model. We can also consider the staircase structure of the ABC. As is clear from~\eqref{ABCgraph}, the leftmost output qubit of each $P_k$ is not affected by any other quantum gate. Hence, it determines how many qubits in the state $\ket{1}$ will flow through the rest of the circuit. It is thus convenient to further split the symmetry reduced maps $P_k^{(r)}$ into two components $P_k^{(i,r)}$, with $i=0,1$,
which make explicit the final state of this qubit
\begin{equation}
P^{(r)}_k  = \left( 
\begin{array}{c} 
P^{(0,r)}_k \\
P^{(1,r)}_k \\
\end{array}
\right) \ .
\label{Pni}
\end{equation}
The depiction of long and short gates is, respectively,
\begin{equation}
\vcenter{\hbox{\includegraphics[width=.75\columnwidth]{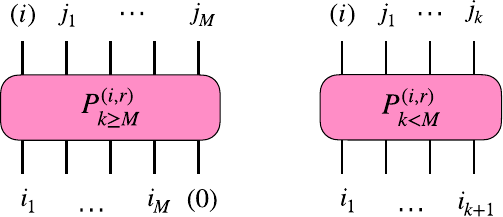}}}
\label{gates3}
\end{equation} 
The input and output configurations to the left map satisfy
\begin{equation}
    \sum_{l=1}^M \, i_l=r \ , \hspace{5mm} \sum_{l=1}^{M} \, j_l=r-i \ .
\end{equation}
An analogous condition holds for the right map
\begin{equation}
    \sum_{l=1}^{k+1} \, i_l=r \ , \hspace{5mm} \sum_{l=1}^{k} \, j_l=r-i \ .
\end{equation}
We recall rounded boxes in~\eqref{gates3} stress that~$P^{(i,r)}_k$ need not be unitary. We have moreover introduced a notation that will serve us along the paper. We denote in parenthesis the state of those qubits which are already explicitly or implicitly described by $P^{(i,r)}_k$, and therefore do not count as input or output to these matrices. This is the case of the leftmost output qubit, associated to the superscript $i$, and the rightmost input qubit of long gates, which is fixed to $\ket{0}$. The main properties of the matrices $P^{(i,r)}_k$ are collected in the following table.
\begin{equation}
\begin{array}{c|c|c|c}
P_k^{(i,r)} & {\rm input} & {\rm output} & {\rm dimensions} \\
\hline 
k<M  & {\cal H}_{r,k+1} & {\cal H}_{r-i,k} & d_{r-i,k} \times d_{r,k+1} \\
\hline 
k \geq M & {\cal H}_{r,M} & {\cal H}_{r-i,M} & d_{r-i,M} \times d_{r,M} \\
\hline
\end{array}
\label{eq:Table2}
\end{equation}

\section{\label{sec:Ansatz}Ansatz for the ABC gates}

Below we propose analytical expressions for the maps~{$P_k^{(i,r)}$} building the ABC gates with arbitrary $M$ and $N$. In order to motivate our ansatz, we start by reviewing the construction of the ABC in the simple case~{$M=1$}, which was derived in~\cite{Sopena22}.

\subsection{One-magnon states}
\label{sec:one}

The Bethe wavefunction~\eqref{Bwf} for one magnon is
\begin{equation}
    |\Psi_N^{(1)}\rangle= \sum_{n=1}^N x^{n-1} |n\rangle \ ,
    \label{Bwf1}
\end{equation}
with $x=e^{i p}$.
When the magnon momentum is real, we obtain a plane wave.  
The previous state is unnormalized. By construction, the ABC should prepare its normalized version
\begin{equation}
    |\Phi_N^{(1)}\rangle= \frac{1}{\sqrt{C_N}} |\Psi_N^{(1)}\rangle \ ,
    \label{Bnorm1}
\end{equation}
where
\begin{equation}
 C_N=  \langle \Psi_N^{(1)}|\Psi_N^{(1)}\rangle = \sum_{n=0}^{N-1} |x|^{2n} \ .
 \label{CN}
\end{equation} 
The circuit~\eqref{ABCgraph} in this simple case only contains two-qubit gates, 
reducing to  
\begin{equation}
\vcenter{\hbox{\includegraphics[width=.8\columnwidth]{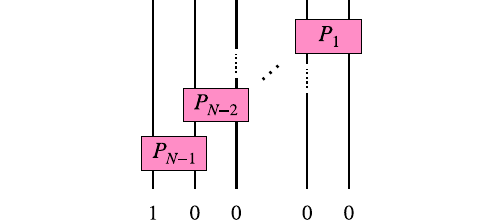}}}  
\label{one1}
\end{equation} 
All gates are of the type denoted in Section~\ref{sec:ABC} 
as long gates, whose rightmost input qubit is always in the state~$\ket{0}$. The decomposition into symmetry protected blocks of their relevant entries is given in~\eqref{bigP}. The only nontrivial block is $P_k^{(1)}$, which can be determined as follows. 

We focus first on the leftmost qubit of the circuit. It is only acted upon by the gate $P_{N-1}$. In order to reproduce the coefficient of the state $\ket{n}=\ket{1}= \ket{10\dots0}$ in~\eqref{Bnorm1}, we need
\begin{equation}
\vcenter{\hbox{\includegraphics[width=.8 \columnwidth]{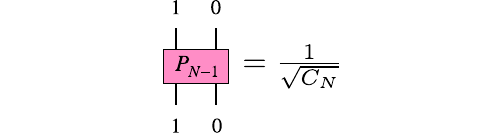}}}  
\label{one2}
\end{equation} 
The second entry of $P^{(1)}_{N-1}$ is then fixed, up to a phase, by unitarity
\begin{equation}
\vcenter{\hbox{\includegraphics[width=0.8\columnwidth]{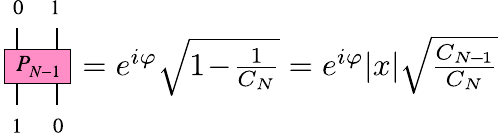}}} 
\label{one3}
\end{equation} 
In the last equality, we have used the recursion relation
\begin{equation}
    C_N=|x|^2 \, C_{N-1}  +1 \ ,
    \label{rec1}
\end{equation}
which is easily derived from~\eqref{CN}. It relates the norms of the wavefunctions supported by $N$ and $N-1$ qubits. 
This is the first of a series of recursion relations with the same property that will be crucial in our construction.
Since~\eqref{one3}
moves the input state $\ket{1}$ one position to the right, it seems natural to adjust the free phase such that it reproduces the plane wave momentum:  $e^{i \varphi}|x|=x$.

We turn now to the second leftmost qubit. It is acted upon by two gates, such that
the coefficient of the state~{$\ket{n}=\ket{2}= \ket{010\dots0}$} in~\eqref{Bnorm1} is reproduced if
\begin{equation}
\vcenter{\hbox{\includegraphics[width=0.8\columnwidth]{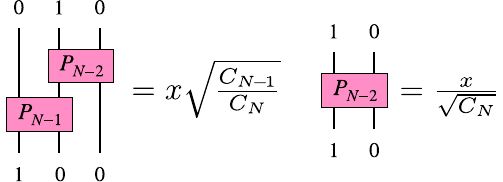}}} \label{one4}
\end{equation} 
Therefore, $P_{N-2}$ must cancel the normalization factor~$C_{N-1}$, such that the correct normalization of the wavefunction of $N$ qubits is restored.
Making the same choice for the free phase, this implies that $P^{(1)}_{N-2}$ is given again by expressions~\eqref{one2} and~\eqref{one3} upon replacing $N$ by $N-1$. We arrive at the following result implied by the recursion relation~\eqref{rec1}. The subcircuit formed by the gates $P_k$
with $k<N-1$ creates a plane wave
of the same momentum $p$  but supported on~$N-1$ qubits. 
Iterating these arguments, we obtain
\begin{equation}
\left( \begin{array}{c} 
P_k^{(0,1)} \\[1mm]
P_k^{(1,1)}  \\
\end{array} 
\right)  = 
    \frac{1}{\sqrt{C_{k+1}} }\begin{pmatrix}
        x \,\sqrt{C_{k}} \\[1mm]
        1
    \end{pmatrix} \ ,
    \label{P1}
\end{equation}
where we have decomposed $P^{(1)}_k$ according to~\eqref{Pni}.
The function $C_k$ encodes the norm of the wavefunction~\eqref{Bwf1} with $N=k$ 
\begin{equation}
    C_k=  \langle \Psi_k^{(1)}|\Psi_k^{(1)}\rangle \ .
    \label{Ck}
\end{equation}
Contrary to the tensor $\mathscr{R}_T$ in~\eqref{ABAcircuit}, which is directly derived from the ABA, the ABC unitaries have an explicit dependence on the gate index $k$.

\subsection{Gate ansatz}

\label{gatean}

We will use the results of the previous section to propose an ansatz describing the general ABC gates, preparing $M$ magnon states. When $M=1$, we only had to consider the normalization of the states emerging from the single $R$ matrices at each step of~\eqref{ABAcircuit}.
In the general case, the distillation of unitaries out of the tensor~$\mathscr{R}_T$ implies a more  involved problem. Namely, the normalization and orthogonalization of the wavefuncions created by the rightmost $k$ tensors of the network~\eqref{ABAcircuit}. Recall that for that partial circuit the input to the leftmost $M$ legs of the lower $\mathscr{R}_T$ can be an arbitrary state.
This means that the single function~\eqref{Ck} has to be promoted to a set of matrices
\begin{equation}
    C_k \, \to \; C^{(r)}_k \ ,
\end{equation}
which should contain the information about norms and overlaps inside each symmetry sector, necessary to turn~\eqref{ABAcircuit} into a valid quantum circuit. We will refer to them as overlap matrices. 
Accordingly, we generalize
\begin{equation}
    \sqrt{C_k} \to A_k^{(r)} \ ,  \;\;\;\;\; \frac{1}{\sqrt{C_{k+1}}} \to B_{k+1}^{(r)}  \ .
\end{equation}
with the matrix $A_k^{(r)}$ denoting the square root of  $C_k^{(r)}$, and $B_k^{(r)}$ being the inverse of $A_k^{(r)}$. Using these definitions, we propose the following ansatz 
\begin{equation}
P_k^{(i,r)} = A_k^{(r-i)} \, \Lambda^{(i,r)} \, B_{k+1}^{(r)} \ .
\label{ansatz}
\end{equation}
While the matrices $A$ and $B$ 
implement the change to an orthonormal basis, 
 $\Lambda^{(i,r)}$ should be responsible for building the coefficients of the Bethe wavefunction~\eqref{Bwf}.

The ansatz~\eqref{ansatz} has a clear MPS inspiration. 
An MPS is an approximation to the multileg tensor in the right hand side (rhs) of~\eqref{gates0} given by the product 
\begin{equation}
\vcenter{\hbox{\includegraphics[width=.81\columnwidth]{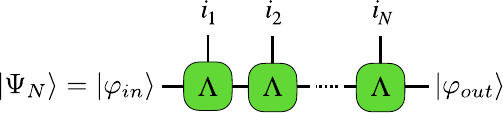}}}  \label{net1}
\end{equation} 
which is defined with the help of 
an auxiliary Hilbert space living on the horizontal links.
The three-leg tensor~$\Lambda$ can be promoted to a matrix 
\begin{equation}
    {\bar \Lambda}: {\cal H}_{\rm aux}\otimes {\cal H}_{\rm phys} \to {\cal H}_{\rm phys}\otimes {\cal H}_{\rm aux} 
    \label{can}
\end{equation}
satisfying ${\bar \Lambda}_{\alpha 0}^{i \beta}= \Lambda^i_{\alpha \beta}$, where $\alpha$ and $\beta$ take values in the auxiliary space and $i$ is a physical index. Using this matrix,~\eqref{net1} admits the circuitlike  representation
\begin{equation}
\vcenter{\hbox{\includegraphics[width=.77\columnwidth]{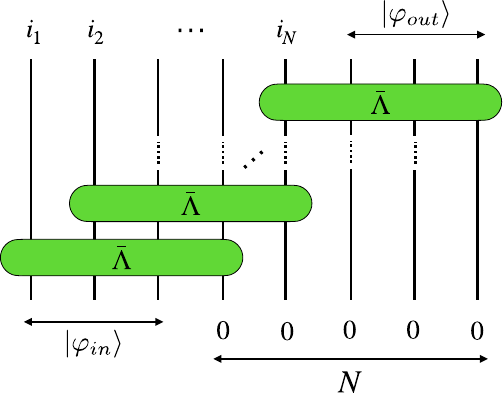}}}  \label{net2}
\end{equation} 
The ABA constructing $M$ magnons states can be very directly interpreted as a MPS network~\cite{Sopena22}. Indeed, the previous figure is equivalent to~\eqref{ABAcircuit}, with~\eqref{can} given by $\mathscr{R}_T$. The auxiliary space in that case is the Hilbert space of~$M$ qubits, and $|\varphi_{in}\rangle=\ket{1_M}$
and $|\varphi_{out}\rangle=\ket{0_M}$. 

The MPS has a large gauge freedom, corresponding to a change of basis in the auxiliary space
\begin{equation}
\vcenter{\hbox{\includegraphics[width=.8\columnwidth]{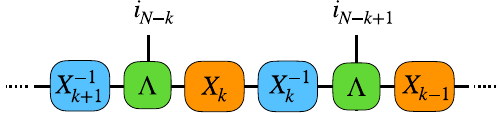}}}  \label{net3}
\end{equation} 
which can vary from link to link. Using this freedom, it is always possible to transform
\begin{equation}
    \Lambda \; \to \; X_{k} \, \Lambda \, X_{k+1}^{-1} \ ,
    \label{MPScan}
\end{equation}
such that the associated matrices~\eqref{can} are unitary and~\eqref{net2} defines a valid quantum circuit.
An MPS with this property is said to be in canonical form~\cite{Cirac09, Orus13, Cirac21}.
The transformation into canonical form was the procedure used in~\cite{Sopena22} to bring the ABA into the ABC. 
The matrices $X_k$ were determined by recursion relations too involved to be analytically solved in the general case. Here we will pursue an ansatz of the same structure but different components, in search of a simplification. Indeed,~\eqref{ansatz} has the form~\eqref{MPScan} with $A$ identified with $X$ and $B$ with its inverse. 
This idea will lead us to an MPS tensor naturally linked to the CBA.

We will focus first on the long gates, whose graphical representation is
\begin{equation}
\vcenter{\hbox{\includegraphics[width=.63\columnwidth]{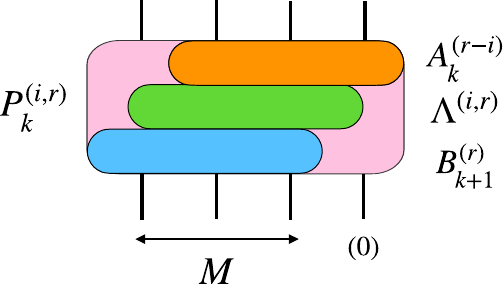}}}  \label{net4}
\end{equation} 
where we used the notation explained in~\eqref{gates3}. All the matrices in the ansatz~\eqref{ansatz} for long gates act on $M$ qubits. Their properties are described in the table below
\begin{equation}
\begin{array}{c|c|c|c}
k \geq M & {\rm input} & {\rm output} & {\rm dimensions} \\
\hline 
B^{(r)}_{k+1}  & {\cal H}_{r,M} & {\cal H}_{r,M} & d_{r,M} \times d_{r,M} \\
\hline 
\Lambda^{(i,r)} & {\cal H}_{r,M} & {\cal H}_{r-i,M} & d_{r-i,M} \times d_{r,M} \\
\hline
A^{(r)}_k  & {\cal H}_{r,M} & {\cal H}_{r,M} & d_{r,M} \times d_{r,M} \\
\hline 
\end{array}
\label{eq:Table3}
\end{equation}
Our ansatz for the short ABC gates, which are crucial for obtaining a deterministic quantum circuit, entails some peculiarities. We postpone their discussion to Section~\ref{sSmallGates}. 

\section{\label{sec:Lambda}The \texorpdfstring{$\Lambda$}{Lamdba} tensors}

\label{scattMat}

Here we present our proposal for the matrices $\Lambda^{(i,r)}$. The simple $M=1$ circuit reviewed in Section~\ref{sec:one} will serve us again as a guide.

\subsection{Diagrammatic rules}

We start by analysing the case when the leftmost output qubit of a long ABC gate is in the state $\ket{0}$, denoted with the superscript $i=0$.
It is instructive to revisit the one magnon result for $P_k^{(0,1)}$ and compare it with the decomposition ~\eqref{ansatz} 
\begin{equation}
\vcenter{\hbox{\includegraphics[width=0.6\columnwidth]{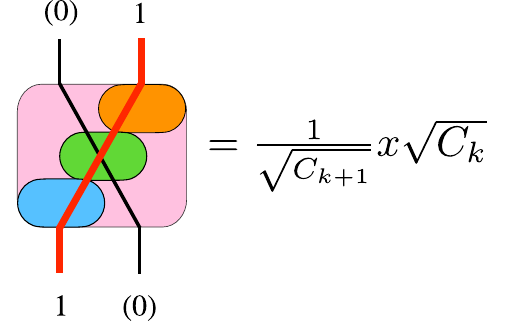}}}  
\label{scatt1}
\end{equation}
In the gate like representation~\eqref{can}, the MPS tensor shifts one site to the right the lines associated to ${\cal H}_{\rm aux}$. For clarity, we represent in parenthesis the qubits belonging to ${\cal H}_{\rm phys}$, on which the matrix $\Lambda^{(i,r)}$ does not act. We also highlight in red the flow of qubits in the state $|1 \rangle$.

From~\eqref{scatt1} we immediately obtain $\Lambda^{(0,1)}=x$, 
which has the appealing interpretation of arising from the displacement of the input $\ket{1}$ one position toward the right.
Extrapolating this intuition to the general case, we propose that~$\Lambda^{(0,r)}$ is a diagonal matrix whose entries are the momentum factors arising from displacing its input state one site to the right. 
This prescription is better understood with a graphical example. 
Let us consider a long gate of an ABC creating a three-magnon state in the $r=2$ symmetry sector. 
The action of $\Lambda^{(0,2)}$ on the input state $|110\rangle$ will be
\begin{equation}
\vcenter{\hbox{\includegraphics[width=0.53\columnwidth]{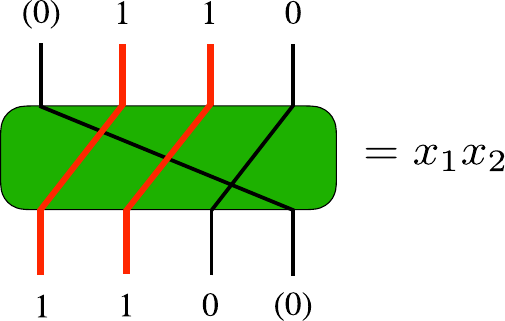}}}  
\label{scatt2}
\end{equation} 
Namely, a factor $x_n$ arises for each red line at the $n$th input position, counted from left to right, where $n$ can take values $1,\dots,M$.

Interestingly,~\eqref{scatt2} hints toward the construction of~$\Lambda^{(1,r)}$.
In this case the leftmost output qubit is by definition in the state $\ket{1}$.
If we follow the same prescription as before, a red line should now traverse the green area from the bottom right to the upper left. But this implies that the rightmost input qubit needs to be in~$\ket{1}$, while by assumption the rightmost input qubit of a long gate is  in $\ket{0}$.
We overcome the difficulty by completing the shift realized by~\eqref{scatt2}  with a previous operation.
It will be 
represented in yellow and implements permutations between the fixed rightmost $\ket{0}$, which we represent graphically as $(0)$, and one of the input qubits in~$\ket{1}$. The rightmost qubit of the output is at $\ket{1}$, and we
denote it by $(1)$.
Taking as example the input state~$|110\rangle$ to $\Lambda^{(1,2)}$, 
 we obtain the following two possible outcomes
\begin{equation}
\vcenter{\hbox{\includegraphics[width=0.73\columnwidth]{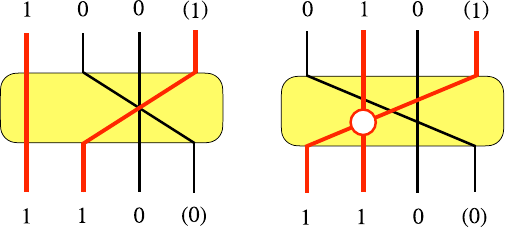}}}  
\label{scatt3}
\end{equation} 
We define
the matrix $\Lambda^{(1,r)}$ as the result of performing first this permutation and then the shift represented by~\eqref{scatt2}. For the same example, the complete action of~$\Lambda^{(1,2)}$ thus is
\begin{equation}
\vcenter{\hbox{\includegraphics[width=0.77\columnwidth]{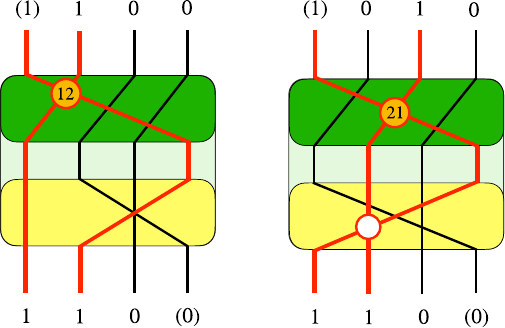}}}  
\label{scatt4}
\end{equation} 
It should be stressed that no shift factor is associated with the red line crossing the green area diagonally from bottom right to upper left. The fact that this line is red, contrary to the situation in~\eqref{scatt2}, is however important.  
When $r>1$ some of the red lines necessarily cross. Every time this happens, we add a scattering contribution which depends on whether the crossing happens in the yellow or green areas
\begin{equation}
\vcenter{\hbox{\includegraphics[width=0.7\columnwidth]{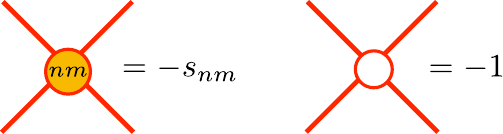}}}  
\label{scatt5}
\end{equation} 
The numbers inside the orange circle indicate which lines cross, with $n$ labelling the left input line and $m$ the right one. The crossings inside the green area carry the dynamical information of the model. The role of those in the yellow area is to ensure the correct antisymmetrization factors.
Collecting all ingredients,~\eqref{scatt4} results into
\begin{equation}
  \Lambda^{(1,2)}|110\rangle=  -s_{12} x_1 |100\rangle +s_{21} x_2 |010\rangle \ .
\end{equation}

Now we translate the previous set of diagrammatic rules into  mathematical expressions.
First, we consider the diagonal matrix~$\Lambda^{(0,r)}$.
In order to characterize the shift toward the right in~\eqref{scatt2}, we introduce collective momentum variables. Associated with an input configuration in the $r$ symmetry sector with qubits in $\ket{1}$ at positions $n_1,\dots,n_r$, we define
\begin{equation}
    x^{(r)}_a=x_{n_1} \ldots x_{n_r} \ .
    \label{y}
\end{equation}
The integer $a=1,\ldots,d_{r,M}$ labels efficiently all possible configurations, as explained in Section~\ref{amap}. See Table~\ref{eq:Table1} for an explicit example of this map. Using the previous variables, we can write
\begin{equation}
   \Lambda^{(0,r)}={\rm diag} \, \big( x^{(r)}_1,\ldots,x^{(r)}_{d_{r,M}}\big)  \ .
   \label{X}
\end{equation}
We define $\Lambda^{(0,0)}=1$, hence $x^{(0)}_1=1$ in the zero-magnon sector. 
Let us move now to $i=1$, where the information about scattering is contained. By definition, $\Lambda^{(1,r)}$ acts as a link between the $r$ and $r-1$ symmetry sectors. 
The transition between the two sectors happens in~\eqref{scatt3}, where an input $1$s is swapped ~with the fixed rightmost $(0)$. This can be described as $a\to a_m$, meaning
\begin{equation}
    (n_1,\ldots , n_r) \to (n_1,\ldots , n_{m-1},n_{m+1},\ldots ,n_r) \ ,
    \label{rel}
\end{equation}
for $m=1,\ldots,r$. The action in the green area of~\eqref{scatt4} is again diagonal, given by~\eqref{X} upon replacing~$r$ with~\mbox{$r-1$}. Finally, we have to add the multiplicative scattering factors~\eqref{scatt5}. Collecting all pieces, we obtain
\begin{equation}
    \Lambda^{(1,r)}_{a_m a}=(-1)^{m+1} s_{n_1 n_m}\ldots s_{n_r n_m} \, x_{a_m}^{(r-1)} \ ,
    \label{S}
\end{equation}
with $a$ and $a_m$ related as in~\eqref{rel}, and $s_{n_m n_m}$ absent. All other entries vanish. 
Notice that the previous expression is consistent with $\Lambda_{1a}^{(1,1)}=1$ with the definition $x^{(0)}_1=1$.

\subsection{From ABC to CBA}

According to the logic of the ansatz~\eqref{ansatz}, the MPS tensor defined by the symmetry restricted maps $\Lambda^{(i,r)}$ 
\begin{equation}
\Lambda^{0}
= 
\left( \begin{array}{cccccc}
1  & 0 &  \cdots & 0 \\
0  & \Lambda^{(0,1)} & \cdots  & 0 \\
\vdots & \vdots &  & \vdots \\[1mm]
0& 0  &  \cdots & \Lambda^{(0,M)}  
\end{array}
\right) \ ,
\label{Lambda1}
\end{equation}
and
\begin{equation}
\Lambda^{1}=
\left( 
\begin{array}{cccccc}
0 & \Lambda^{(1,1)}  & \cdots  & 0 \\
\vdots & \vdots & & \vdots\\
0 & 0 & \cdots & \Lambda^{(1,M)}  \\
0 & 0  & \cdots & 0 \\
\end{array}
\right) \ ,
\label{Lambda2}
\end{equation}

\vspace{2mm}
\noindent
should reproduce the Bethe wavefunctions~\eqref{Bwf}. We will prove in Appendix~\ref{appBTN} that this is indeed the case. Here we illustrate how the component $\ket{n_1n_2}=\ket{13}$ of the two-magnon state~\eqref{B2wf} is reproduced by our diagrammatic rules. For that, it is enough to analyse the action of the first three $\Lambda$ matrices  
\begin{equation}
\vcenter{\hbox{\includegraphics[width=0.88\columnwidth]{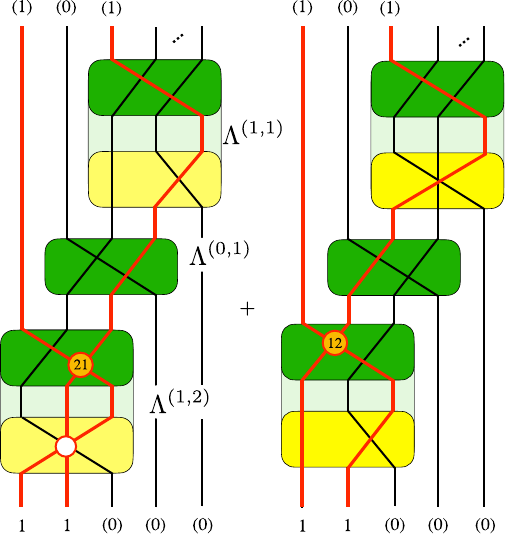}}} 
\label{scatt6}
\end{equation} 
We correctly obtain
\begin{equation}
    \langle 13 | \Psi^{(2)}_N \rangle= s_{21} x_2^2-s_{12}x_1^2 \ .
    \label{ex13}
\end{equation}

We stress that our diagrammatic rules exactly prepare the Bethe wavefunction~\eqref{Bwf}. Namely, no proportionality factor needs to be included, contrary to the state~\eqref{BABA} constructed by the ABA. It should be noted that determining the proportionality factor between the states created by the ABA and the CBA is a complicated problem.

Thus, we can conclude that 
~\eqref{X} and~\eqref{S} provide the MPS tensor naturally associated with the CBA.
Although the map between both tensors must be a transformation of the form~\eqref{net3}, it is necessarily nontrivial. The ABA tensor for multimagnon states is constructed out of a product of $R$ matrices, each of them depending on an individual magnon momenta. On the contrary, the $\Lambda$ defined above only depends on the scattering amplitude of the model, which is a function of pairs of magnon momenta.  

\section{\label{sec:A_B}The \texorpdfstring{$A$}{A} and \texorpdfstring{$B$}{B}  matrices}

\label{overlapMat}

The MPS tensor associated to the CBA derived above and that of the ABA do not lead directly to valid quantum gates. They need to be brought into canonical form using the freedom~\eqref{net3}. In this and the next section we will determine the change of basis needed to
transform~\eqref{X} and~\eqref{S} into the quantum gates $P_k$.

\subsection{Overlap matrices}

In the search for the required transformation, let us consider the subcircuit formed by the first $k-1$ gates of the ABC, highlighted in yellow below
\begin{equation}
\vcenter{\hbox{\includegraphics[width=0.71\columnwidth]{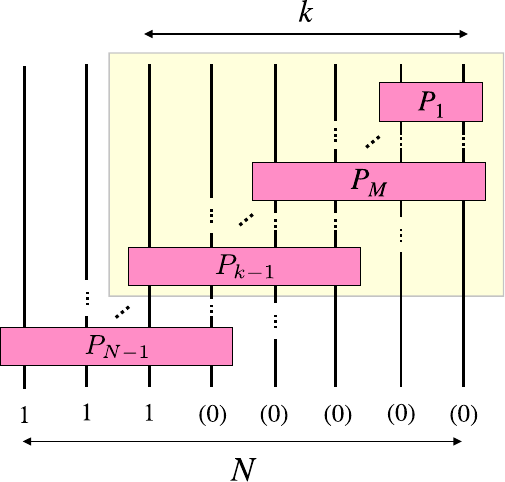}}\vspace{-2mm}}  
\label{overlap1}
\end{equation} 
We assume that $k>M$, such that $P_{k-1}$ is a long gate. 
While the input configuration to $P_{N-1}$ is fixed to~\mbox{$\ket{1_M} \otimes \ket{0}$}, that of a general long gate can have the first $M$ qubits in an arbitrary state.
A basis for the input states to $P_{k-1}$ is thus $|n_1\ldots n_r\rangle$, where the number of qubits in $\ket{1}$ can take any value $r\leq M$ and~\mbox{$n_r \leq M$} assures that the rightmost qubit is in $\ket{0}$.
For each basis element, the subcircuit highlighted in yellow will output a different state
supported by $k$ qubits
\begin{equation}
   |n_1 \dots n_r\rangle \to |\Phi^{(r)}_{k,a}\rangle \ .
   \label{functions}
\end{equation}
The integer $a=1,\ldots,d_{r,M}$ is related to $(n_1,\ldots,n_r)$ as explained in Section~\ref{amap}. Since the circuit is unitary, all these states are orthonormal
\begin{equation}
\label{orthonormal}
    \langle \Phi^{(r)}_{k,a}|\Phi^{(r)}_{k,b} \rangle = \delta_{ab} \ .
\end{equation}

The MPS based on the tensor $\Lambda$ derived in the previous section builds the Bethe wavefunctions~\eqref{Bwf}. Equivalently, the probabilistic circuitlike network~\eqref{net2}, with~${\bar \Lambda}$ associated to $\Lambda$ as explained in~\eqref{can}, outputs Bethe wavefunctions when the $M$ rightmost auxiliary qubits are projected into the reference state $\ket{0}$. The ABC is, however, a deterministic circuit which needs no auxiliary qubits. Based on the ansatz~\eqref{ansatz}, we have interpreted the ABC as the unitary version of~\eqref{net2}.
The mismatch between the number of qubits involved in both realisations of~\eqref{Bwf} renders their correspondence nontrivial. 
This issue will be addressed in the next section. 
Here it is enough to notice that  the subcircuit in~\eqref{overlap1} must be equivalent for consistency to
\begin{equation}
\vcenter{\hbox{\includegraphics[width=0.72\columnwidth]{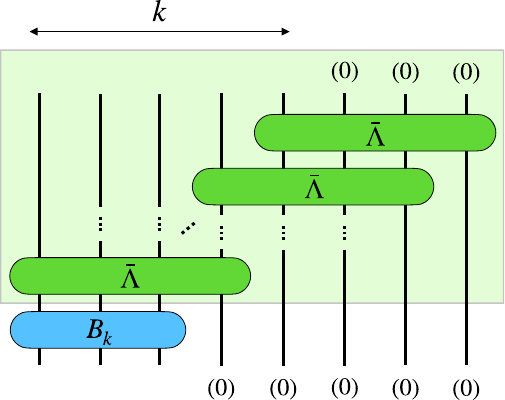}}} 
\label{overlap2}
\end{equation}  

The matrices ${\bar \Lambda}$ in the green area create $k$ qubit states~$|\Psi_{k,a}^{(r)}\rangle$, with the same conventions as in~\eqref{functions}. Since ${\bar \Lambda}$ is not unitary, these states are not orthogonal and they are not normalized. We define a matrix of overlaps in each symmetry sector, whose entries are
\begin{equation}
    C_{k,ab}^{(r)}=\langle \Psi^{(r)}_{k,a}|\Psi^{(r)}_{k,b} \rangle \ .
    \label{inner}
\end{equation}
This is known in mathematical terms as a Gram matrix.
To ensure that~\eqref{overlap2} coincides with the highlighted area of~\eqref{overlap1}, we need $B$ to transform the non orthogonal set into an orthonormal set, that is
\begin{equation}
    B^{(r)\dagger}_k C_k^{(r)} B^{(r)}_k =\mathbb{1}_{d_{r,M}} \ .
    \label{BCB}
\end{equation}
This relation can only be satisfied if the overlap matrix is positive definite. By definition it is also Hermitian.
A matrix with these properties can be subject to the so called Cholesky decomposition
\begin{equation}
    C_k^{(r)}=A^{(r)\dagger}_k \, A^{(r)}_k  \ .
    \label{Cholesky}
\end{equation}
The matrix $A$ becomes unique if we chose it to be upper triangular and with positive diagonal entries. Relation~\eqref{BCB} fixes $B$ up to a unitary transformation. 
Substituting the Cholesky formula into~\eqref{BCB}, we fix that freedom by requiring that the matrices $A$ and $B$ are inverse of each other, as we mentioned above~\eqref{ansatz}. With this choice, $B$ will also be upper triangular. Explicit formulae for $A$ and $B$ in terms of minors of the overlap matrices can be found in Appendix~\ref{appAB} . These matrices, together with~\eqref{X} and~\eqref{S}, complete the ansatz~\eqref{ansatz} for the long gates.

\subsection{Partial Bethe wavefunctions}

The previous construction would remain rather formal unless we determine the states 
$|\Psi_{k,a}^{(r)}\rangle$. 
These states turn out to be again Bethe wavefunctions, with momenta drawn from the magnon momenta defining the state prepared by the complete quantum circuit. The subset of momenta associated to $|\Psi_{k,a}^{(r)}\rangle$ is determined by the input state to the shaded region in~\eqref{overlap2} as follows
\begin{equation}
    |n_1 \dots n_r\rangle \to (p_{n_1},\dots,p_{n_r}) \ ,
    \label{momenta}
\end{equation}
with $a$ related to $(n_1,\ldots,n_r)$.
This assignment is proven in Appendix~\ref{appBTN}. Here we 
will limit ourselves to give an example of how it  arises.

Let us consider an ABC creating a Bethe wavefunction describing three magnons, and  input to the shaded region of~\eqref{overlap2} the state $|n_1n_2\rangle=\ket{12}$ from the $r=2$ symmetry sector. The coefficient multiplying the basis element $\ket{13}$ of the output is obtained by adding the contributions of the graphs
\begin{equation}
\vcenter{\hbox{\includegraphics[width=.88\columnwidth]{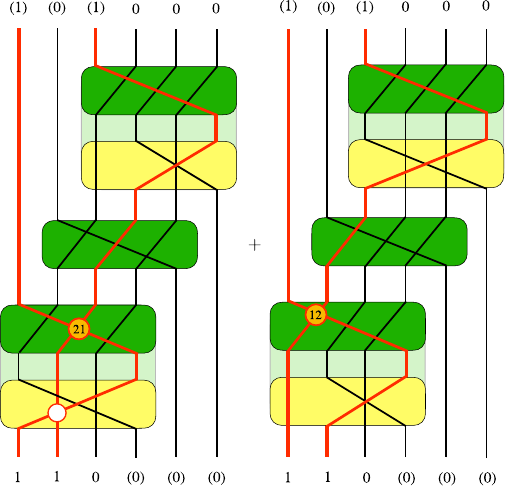}}}  
\label{overlap3}
\end{equation} 
Using the rules of Section~\ref{scattMat}, we obtain
\begin{equation}
    \langle 13 | \Psi^{(2)}_{k,1} \rangle= s_{21} x_2^2-s_{12}x_1^2 \ ,
\end{equation}
This result coincides with~\eqref{ex13}, which emerges from the ABC designed to build a two-magnon Bethe wavefunction~\eqref{B2wf} with momenta $p_1$ and $p_2$.
If the input state is instead $|n_1n_2\rangle=\ket{13}$,
the coefficient of the same output element $\ket{13}$ will be
\begin{equation}
    \langle 13 | \Psi^{(2)}_{k,2} \rangle= s_{31} x_3^2-s_{13}x_1^2 \ ,
    \label{pag}
\end{equation}
which is now consistent with a two-magnon output state of momenta $p_1$ and $p_3$. As before,~\eqref{pag} arises from the contributions of two graphs.
These are 
\begin{equation}
\vcenter{\hbox{\includegraphics[width=.88\columnwidth]{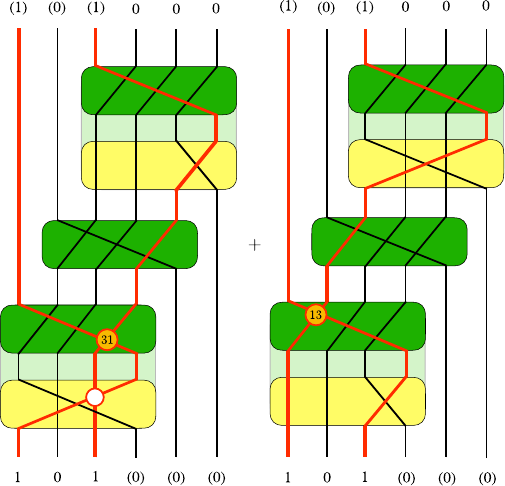}}}  
\label{overlap4}
\end{equation} 

Notice that the states $|\Phi^{(r)}_{k,a}\rangle$ resulting from the subcircuit in~\eqref{overlap1} are in general not Bethe wavefunctions. This is only the case for $r=M$, since there is only one state in that symmetry sector. Hence 
\begin{equation}
    |\Phi^{(M)}_{k,1}\rangle= \frac{1}{ \sqrt{\langle\Psi^{(M)}_{k,1}|\Psi^{(M)}_{k,1}\rangle}} \,|\Psi^{(M)}_{k,1}\rangle \ ,
\end{equation}
which generalizes the analogous statement for $M=1$, obtained in Section~\ref{sec:one}. Namely, 
when the input to the shaded circuit in~\eqref{overlap1} is $\ket{1_M}\otimes \ket{0}$, it outputs the same Bethe wavefunction prepared by the complete circuit, but supported by $k$ instead of $N$ qubits. 

\section{\label{sec:short_gates}The short gates}

\label{sSmallGates}

Long and short gates differ in the number of qubits on which they act. The latter act upon 
a number of qubits that grows with their gate index as $k+1$ until $k=M$ is reached. This property is essential for the ABC to define a deterministic circuit. 
A graphical representation of~\eqref{ansatz} applied to short gates is
\begin{equation}
\vcenter{\hbox{\includegraphics[width=0.63\columnwidth]{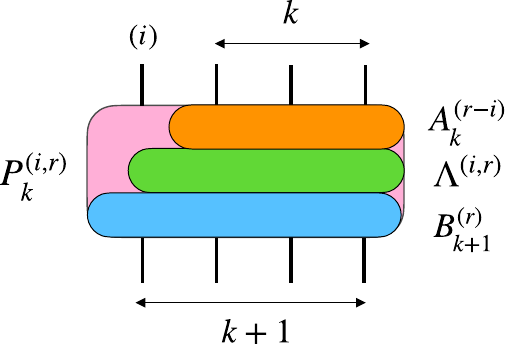}}} 
\label{small1}
\end{equation} 
Some adjustments in the construction of $\Lambda$ and the unitarization matrices $A$ and $B$ with respect to previous sections are clearly necessary. 

\subsection{A ghost leg for \texorpdfstring{$\Lambda$}{Lambda}}

\label{ghost}

The dependence of $\Lambda$ on $k$ just comes into play through the number of qubits on which it acts. For this reason, we will keep the previous notation and do not add to it a gate subscript.
We want the matrix $\Lambda$ of short gates for consistency to be based on the same diagrammatic rules presented in Section~\ref{scattMat}. But this raises  a caveat. 
The fixed input $\ket{0}$ of the long gates's rightmost qubit played an essential role. 
The input state to the short gates is instead unconstrained, making unclear how to apply the diagrammatic rules. 
We solve this problem by adding a rightmost $\ket{0}$ as a ghost input, namely, an input which does not correspond to a real qubit. It will be represented graphically by a $\bullet$. 
We consider first the matrices $\Lambda^{(0,r)}$, resorting again to an example. With the help of the ghost input, the action on the state $\ket{101}$ of the matrix $\Lambda^{(0,2)}$ as part of the short gate $P_2$ is given by
\begin{equation}
\vcenter{\hbox{\includegraphics[width=0.52\columnwidth]{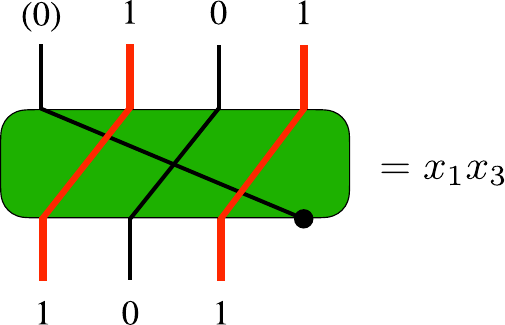}}}  
\label{small2}
\end{equation} 
The same trick  applies straightforwardly to the matrices~$\Lambda^{(1,r)}$. For instance, the action of $\Lambda^{(1,2)}$ belonging to the short gate $P_1$ on $\ket{11}$ leads to 
\begin{equation}
\vcenter{\hbox{\includegraphics[width=0.66\columnwidth]{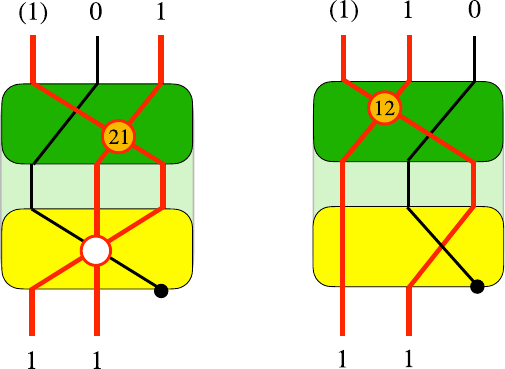}}}  
\label{small3}
\end{equation} 

Hence, the new $\Lambda^{(i,r)}$ provide a map
between input and output configurations of $k+1$ physical qubits. As a result, the matrix $A$ for short gates
must connect $k+1$ with $k$ qubit configurations. This means that it extinguishes the ghost leg introduced by $\Lambda$
\begin{equation}
\vcenter{\hbox{\includegraphics[width=0.37\columnwidth]{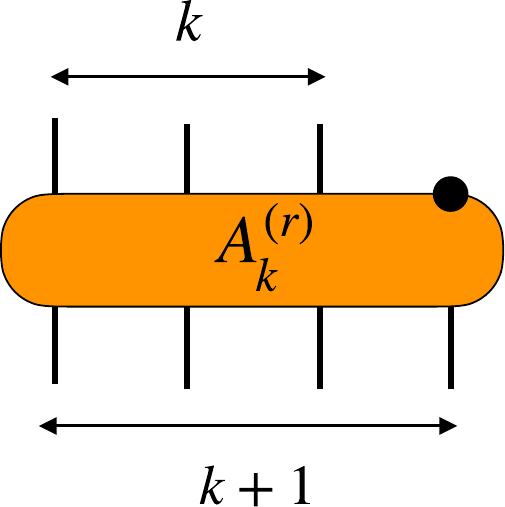}}}  
\label{small4}
\end{equation} 
We represent this again with a $\bullet$, which we define to always absorb a $\ket{0}$.
Let us give an example of how these new rules fit together in the decomposition~\eqref{small1}, using a graph 
associated to the short gate $P_2$ 
\begin{equation}
\vcenter{\hbox{\includegraphics[width=0.48\columnwidth]{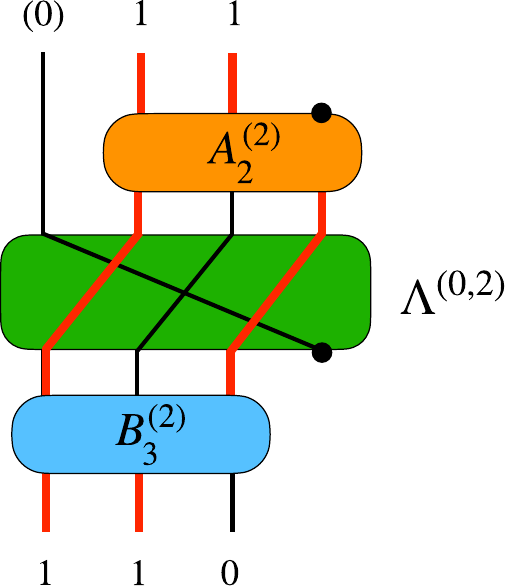}}} 
\label{small5}
\end{equation} 

The dimensions of the matrices $A$, $B$ and $\Lambda$ building the short gates are summarized in the Table below 
\begin{equation}
\begin{array}{c|c|c|c}
k < M & {\rm input} & {\rm output} & {\rm dimensions} \\
\hline 
B^{(r)}_{k+1}  & {\cal H}_{r,k+1} & {\cal H}_{r,k+1} & d_{r,k+1} \times d_{r,k+1} \\
\hline 
\Lambda^{(i,r)} & {\cal H}_{r,k+1} & {\cal H}_{r-i,k+1} & d_{r-i,k+1} \times d_{r,k+1} \\
\hline
A^{(r)}_k  & {\cal H}_{r,k+1} & {\cal H}_{r,k} & d_{r,k} \times d_{r,k+1} \\
\hline 
\end{array}
\label{Tableshort}
\end{equation}
The mathematical expressions for $\Lambda$ are still given by~\eqref{X} and~\eqref{S},
after substituting $M$ by $k+1$~\footnote{
Although the dimensions of $\Lambda$ depend on $k$, we keep this dependence implicit and do not add the corresponding subscript. In this way we stress that the same principle defines $\Lambda$ for long and short gates.
}. 
The unifying definition of $\Lambda$ for long and short gates has however the consequence that~\eqref{small1} cannot be interpreted as a transformation into canonical form. 
Since~$A$ is not a square matrix, it does not correspond to a change of basis. Such an interpretation could be restored with a different choice of the elements in the decomposition~\eqref{small1}, but it would come at the expense of obscuring their meaning, therefore this will not be  pursued here.
 
\subsection{Small overlap matrices}

To determine the matrices $A$ and $B$ of the short gates, we  follow the logic outlined in Section~\ref{overlapMat}.
The matrix $B$ should describe a change of basis that orthonormalizes the states prepared by a part of the ABC. In analogy with the shaded area of~\eqref{overlap2}, this part is
\begin{equation}
\vcenter{\hbox{\includegraphics[width=0.48\columnwidth]{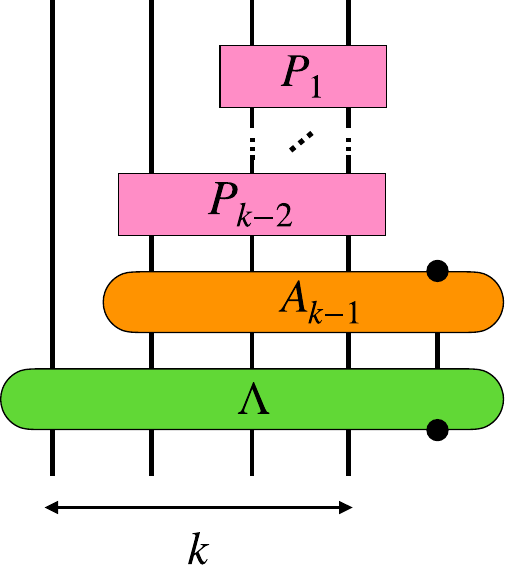}}} 
\label{small6}
\end{equation} 
We have labelled the green gate with the three-leg tensor $\Lambda$ instead of its matrix version $\bar \Lambda$ to emphasize that the ghost leg implies no additional qubit. Carrying on the parallel with long gates, 
when a basis state~{$|n_1\ldots n_r\rangle$} enters~\eqref{small6}, the output should be a $k$ qubit Bethe wavefunction with momenta drawn from the subset $p_1,\ldots,p_{k}$ according to the rule~\eqref{momenta}. Unfortunately we were not able to fully prove this statement. 
In spite of that, we take it as valid and determine $B$ by the requirement 
\begin{equation}
    B^{(r)\dagger}_{k}\, C_{k}^{(r)} \, B^{(r)}_{k} =\mathbb{1}_{d_{r,k}} \ .
    \label{BCB2}
\end{equation}
The overlap matrix $C$ has dimension $d_{r,k} \times d_{r,k}$ and is defined as in~\eqref{inner}. 

Setting $A^{(r)}_k$ to be the inverse of $B^{(r)}_k$ is not compatible with the dimensions described in Table~\ref{Tableshort}. 
The matrix~{$A_k^{(r)}$} has been taken to bridge between configurations of~{$k+1$} and $k$ qubits. Guided by this, we build a new map from $k+1$ qubit 
configurations into states supported by $k$ qubits
\begin{equation}
    |n_1 \dots n_r \rangle \to |{\widehat \Psi}^{(r)}_{k,a}\rangle \ ,
    \label{newstates}
\end{equation}
where $r\leq k$ by $U(1)$ conservation.
The magnon momenta are selected  now from the enlarged set ~{$p_1,\ldots,p_{k+1}$} according again to~\eqref{momenta}. The overlap matrix associated with~\eqref{newstates} has dimension $d_{r,k+1} \times d_{r,k+1}$, and by construction contains the previous one.

Since ${\cal H}_{r,k}$ 
has dimension $d_{r,k}$, not all states $|{\widehat \Psi}^{(r)}_{k,a}\rangle$ can be linearly independent.  As a result, their overlap matrix $\widehat{C}_k^{(r)}$ will be positive \textit{semidefinite}. The Cholesky decomposition of such a matrix  still exists
\begin{equation}
    {\widehat C}_k^{(r)}=A^{(r)\dagger}_k \, A^{(r)}_k  \ , \ 
    \label{Cholesky2}
\end{equation}
but, contrary to~\eqref{Cholesky}, $A_k^{(r)}$ is now upper ${\it rectangular}$. Its number of rows is determined by the rank of the matrix that we want to decompose, which in our case is $d_{r,k}$. 
Hence, it has the right dimensions to constitute our guess for the corresponding matrix in~\eqref{small1}.

\subsection{The product matrix \texorpdfstring{$L$}{L}} 

Although the product of $A$ and $B$ for long gates is the identity, this product for short gates defines a relevant new matrix
\begin{equation}
\vcenter{\hbox{\includegraphics[width=0.82\columnwidth]{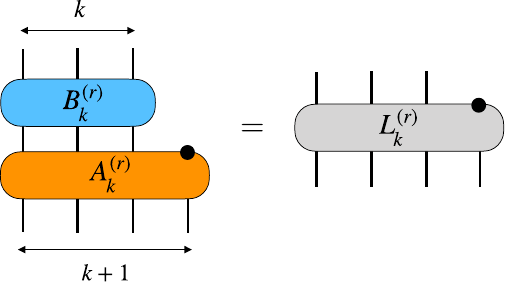}}}  
\label{small7}
\end{equation} 
The matrix $L$ has  dimension $d_{r,k} \!\times  d_{r,k+1}$ with $r\!\leq k$. 
Given that $\widehat C$ contains $C$, we can fix $B$ to be the inverse of the upper {\it triangular} block of $A$. Appendix~\ref{appAB} then shows that
\begin{equation}
     L^{(r)}_{k,ab} = \frac{{\rm det} \; C^{(r)}_{k,a \to b}}{{\rm det} \;  C^{(r)}_{k}} \ ,
     \label{Lmatrix1}
\end{equation}
which is easily seen to 
reduce to the identity when the  rightmost leg is in $\ket{0}$. The
notation $a\to b$ means that the $a$th row of $C$ is substituted by the $b$th row of $\widehat C$.

Using the matrix $L$,~\eqref{small6}
can be recast as
\begin{equation}
\vcenter{\hbox{\includegraphics[width=0.43\columnwidth]{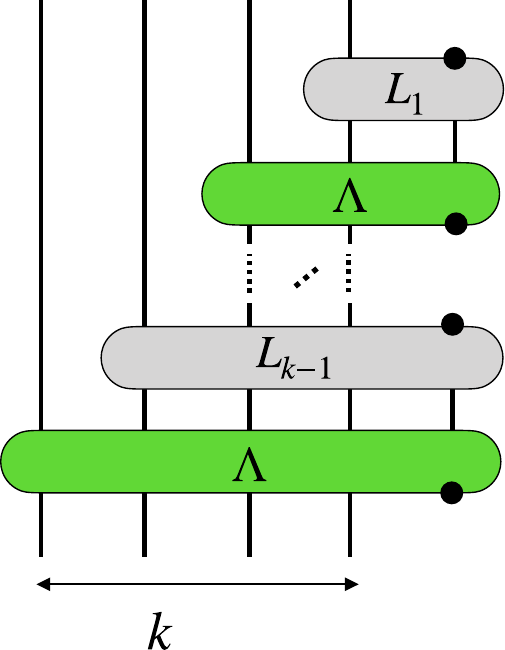}}}  
\label{small8}
\end{equation} 
In the derivation of the matrices $A$ and $B$, we have assumed that this network builds Bethe wavefunctions. In Appendix~\ref{appBTN} on the other hand, these states are shown to be created by the probabilistic circuitlike network
\begin{equation}
\vcenter{\hbox{\includegraphics[width=0.69\columnwidth]{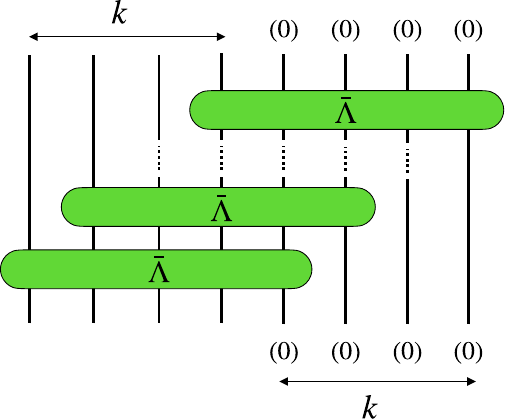}}}  \label{small9}
\end{equation} 
Hence the correctness of the short gate construction relies upon the equivalence of~\eqref{small8} and~\eqref{small9}. In particular,  it completes the proof that the $M$ magnon ABC on $N$ qubits builds the desired Bethe wavefunction when $k=M$. The different dimensions of the green gates in~\eqref{small8} and~\eqref{small9} however render a general demonstration of the equivalence difficult. We have checked that it holds up to~\mbox{$k=5$} using computer algebra software, and tested it numerically up to $k=10$ with \texttt{Qibo}~\cite{qibo_paper}, an open source library to simulate quantum circuits. The programs to reproduce these results are available in~\cite{code_abc}.

\subsection{An example} 

In order to better understand the role of the matrices~$L$ in the previous equivalence, we will analyse the simple case $k=3$. 
Disregarding the first green gate of each side because they have the same action on physical qubits, it reduces to prove
\begin{equation}
\vcenter{\hbox{\includegraphics[width=0.83\columnwidth]{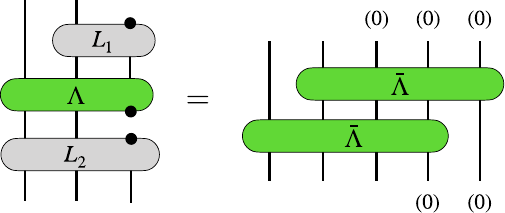}}}  
\label{small10}
\end{equation} 
We will need the explicit expressions of $L_{1,2}^{(r)}$. Using~\eqref{Lmatrix1}, they are given by
\begin{eqnarray}
   L_1^{(1)}&\!\!\!\!=\!\!\!&(1,1) \ , \;\;\;\;\;\;L_2^{(1)} \,=\; \begin{pmatrix}
        \; 1 & 0 & \frac{x_3-x_2}{x_1 -x_2} \label{L21}\\[2mm]
        0 & 1 & \frac{x_1-x_3}{x_1 -x_2} \end{pmatrix} \ , \\[3mm]
       L_2^{(2)}&=& 
            \Big(\, 1 ,\,\frac{s_{31} x_3-s_{13} x_1}{s_{21} x_2 -s_{12}x_1} , \,\frac{s_{32} x_3-s_{23} x_2}{s_{21} x_2 -s_{12}x_1}\, \Big)
         \ . \;\;\;\;\;\;\; \label{L22}
\end{eqnarray}

When the rightmost input qubit to the left hand side (LHS) of~\eqref{small10} 
is in $\ket{0}$, the action of $L_2$ becomes trivial. We thus rather focus on input configurations where that qubit is in $\ket{1}$.
The input state $|001\rangle$ leads to two possible outputs, $|01\rangle$ and $|10\rangle$. Two different braids connect~$|001\rangle$
with $|01\rangle$ 
\begin{equation}
\vcenter{\hbox{\includegraphics[width=0.64\columnwidth]{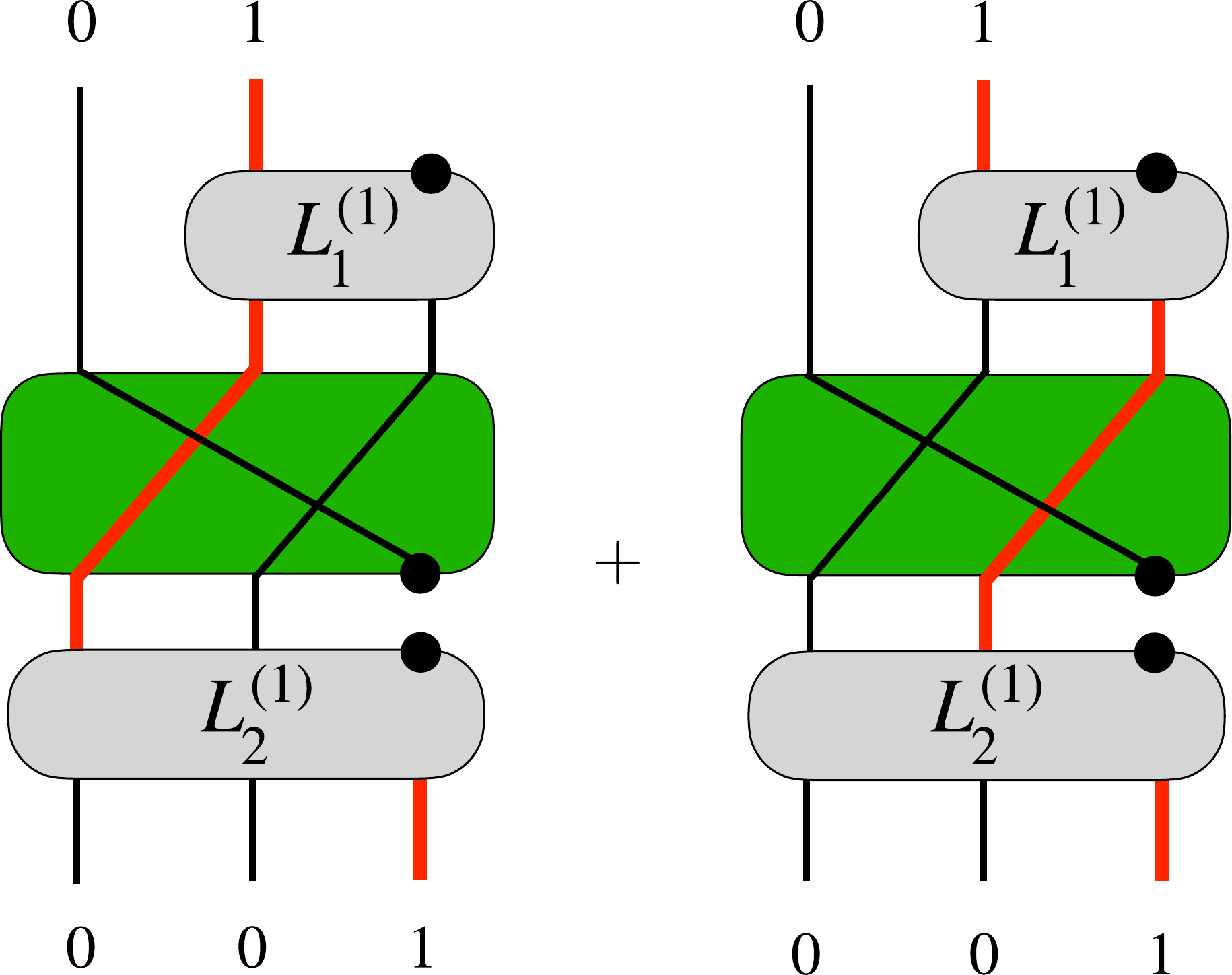}}}  
\label{small12}
\end{equation} 
Substituting the appropriate entries of $L_{1,2}^{(1)}$, the contribution of these graphs leads to 
\begin{equation}
   x_1\frac{x_3-x_2}{x_1 -x_2}+ x_2\frac{x_1-x_3}{x_1 -x_2}=x_3 \ .
\end{equation}
This correctly reproduces the result of the single path allowed by the rhs of~\eqref{small10} to connect those input and out states
\begin{equation}
\vcenter{\hbox{\includegraphics[width=0.57\columnwidth]{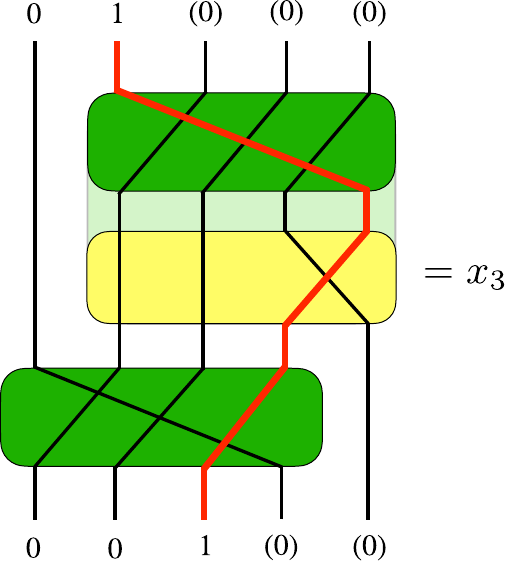}}}  
\label{small13}
\end{equation} 

The red line in the single path above visits auxiliary qubits which do not have a counterpart in~\eqref{small12}. 
The rightmost line entering $L_2$ in~\eqref{small12} is a continuation of the ghost leg created by the previous matrix ${\bar \Lambda}$. When that leg carries a $\ket{1}$, and before extinguishing it, the mission of $L_2$ is to bring back the $\ket{1}$ into the flow of physical qubits. This is the crucial point that makes the equivalence in~\eqref{small10} possible. 
An analogous analysis holds for the transition between $\ket{001}$ and $|10\rangle$. The graphs 
\begin{equation}
\vcenter{\hbox{\includegraphics[width=0.6\columnwidth]{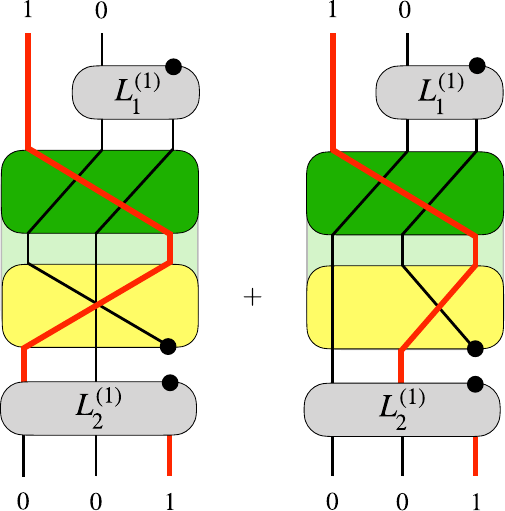}}}  
\label{small14}
\end{equation} 
contribute now
\begin{equation}
   \frac{x_3-x_2}{x_1 -x_2}+ \frac{x_1-x_3}{x_1 -x_2}=1 \ ,
\end{equation}
in agreement with
\begin{equation}
\vcenter{\hbox{\includegraphics[width=0.49\columnwidth]{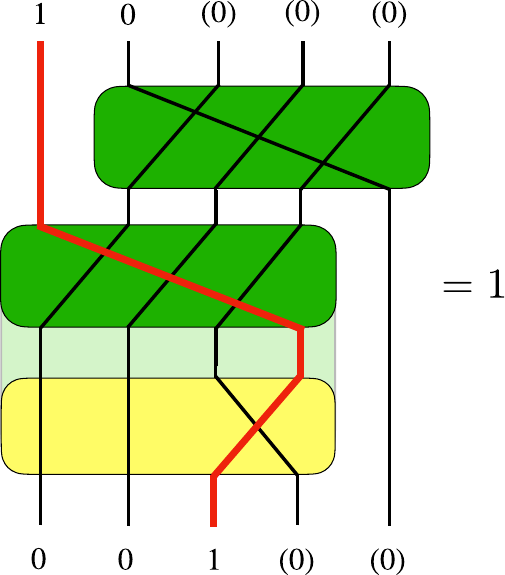}}}  
\label{small15}
\end{equation} 
Finally, we consider input states
$|101\rangle$ and $|011\rangle$. Both of them connect with the output  $|11\rangle$. We will only analyse the input $|101\rangle$ because, in this case, the other does not bring any new insight. The LHS of~\eqref{small10} gives raise again to two paths 
\begin{equation}
\vcenter{\hbox{\includegraphics[width=0.62\columnwidth]{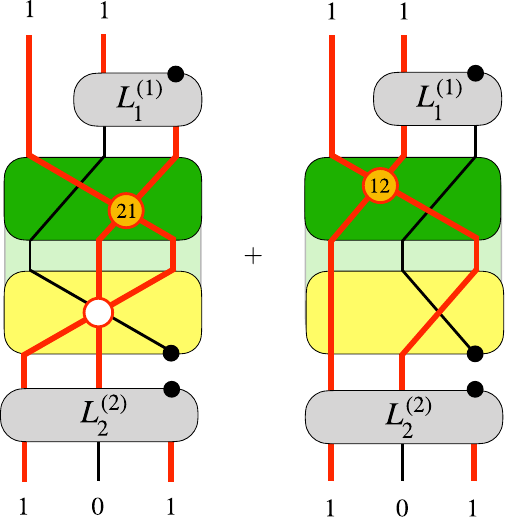}}}  
\label{small16}
\end{equation} 
The matrix $L_2$ contributes equally to both, obtaining
\begin{equation}
     \frac{s_{31} x_3-s_{13} x_1}{s_{21} x_2 -s_{12}x_1} (s_{21} x_2-s_{12}x_1) \ .
     \label{agree}
\end{equation}
This time also the rhs of~\eqref{small10} leads to two paths
\begin{equation}
\vcenter{\hbox{\includegraphics[width=0.85\columnwidth]{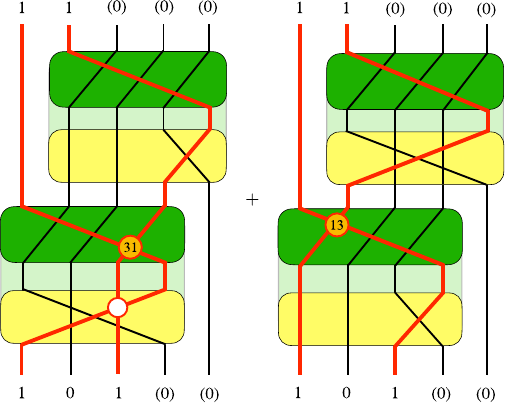}}}  
\label{small17}
\end{equation} 
whose contribution agrees with~\eqref{agree}.
It is tempting to 
associate the first and second graphs in~\eqref{small16} separately with the first and second graphs in~\eqref{small17}. Their separate contributions do not coincide however. Agreement is only found between their sums, providing one more witness to the intricacy of the short gate construction.  

\section{\label{sec:Unitarity}Unitarity}

We have completed our proposal for the ABC gates and given strong arguments that they prepare the desired Bethe wavefunctions.
The last and fundamental requirement to prove is unitarity.

We address again first the long gates. For the purpose of the ABC, it was only necessary to specify how they act when the rightmost input qubit is in $\ket{0}$. The condition for $P_k\ket{0}$ to be promoted to a unitary matrix acting on $M+1$ qubits, is
\begin{equation}
\langle 0 | P_k^\dagger P_k^{\vphantom{\dagger}} |0 \rangle = {\mathbb 1}_{2^{M}} \ ,
\end{equation}
for $k\geq M$. There is no unique way to realize this embedding and thus can be adjusted to the quantum architecture at our disposal. Inside each symmetry sector, the previous condition becomes
\begin{equation}
\label{unitarity}
    P_k^{(0,r)\dagger}P_k^{(0,r)}+P_k^{(1,r)\dagger}P_k^{(1,r)}={\mathbb 1}_{d_{r,M}} \ .
\end{equation}
We substitute the ansatz~\eqref{ansatz} for the ABC gates, and eliminate the resulting matrices $A$ using the Cholesky formula~\eqref{Cholesky}. Unitarity translates then into
\begin{equation}
    B_{k+1}^{(r)\dagger}\! \!\left( \!\sum_{\;i=0,1} \Lambda^{(i,r)\dagger} 
    C^{(r-i)}_{k}\Lambda^{(i,r)} \!\right) \!B_{k+1}^{(r)} ={\mathbb 1}_{d_{r,M}} \ .
    \label{unit}
\end{equation}

The following figure gives a representation of $C_{k+1}^{(r)}$, provided the input and output states are $M$ qubit configurations $\ket{n_1\dots n_r}$ in the $r$ symmetry sector
\begin{equation}
\vcenter{\hbox{\includegraphics[width=0.7\columnwidth]{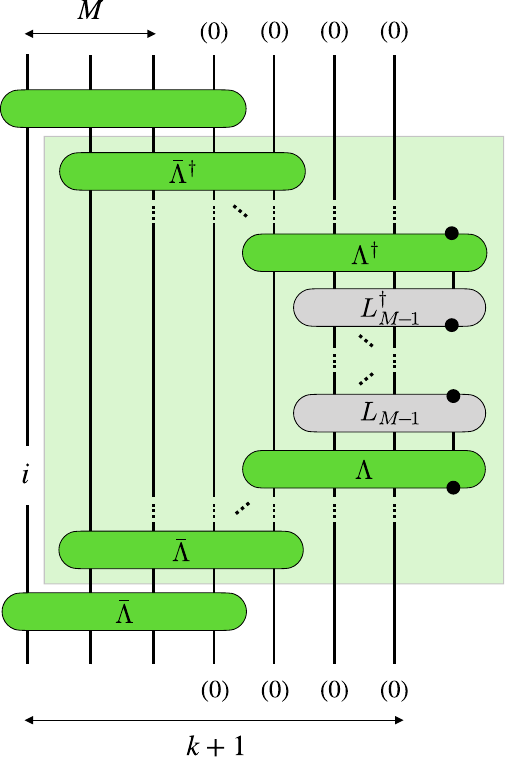}}} 
\label{unitarity1}
\end{equation} 
and $i$ runs on the two values $0$ and $1$. The network in the shaded area builds then $C_{k}^{(r-i)}$. Hence~\eqref{unitarity1} leads to the recursion relations
\begin{equation}
    C^{(r)}_{k+1}=\sum_{i=0,1} \Lambda^{(i,r)\dagger} 
    C^{(r-i)}_{k}\Lambda^{(i,r)} \ ,
\label{reclong}
\end{equation}
with $C^{(0)}_k=1$.
These relations summarize the structure that made possible to find an explicit solution for the ABC gates. They turn out to also guarantee their unitarity. Indeed, inserting it into~\eqref{unit} we recover the change of basis relation that defines $B$.
When \mbox{$M=1$,~\eqref{reclong}} reduces to the simple recursion relation~\eqref{rec1}.

The treatment of the short gates is analogous. The main difference lies in the Cholesky decomposition, since we should use formula~\eqref{Cholesky2} instead of~\eqref{Cholesky}, obtaining
\begin{equation}
   B_{k+1}^{(r)\dagger}\!\! \left( \!\sum_{\;i=0,1} \Lambda^{(i,r)\dagger} 
    {\widehat C}^{(r-i)}_{k}\Lambda^{(i,r)}\! \right) \!B_{k+1}^{(r)} ={\mathbb 1}_{d_{r,k+1}} ,
    \label{unit2}
\end{equation}
for $k<M$. As a consequence, the overlap matrix $C$ in~\eqref{unit} is replaced by
${\widehat C}$, defined in terms of the enlarged set of states~\eqref{newstates}. From the definitions of the matrices $A$ and $B$ for short gates, it is easy to see that both overlap matrices are related by
\begin{equation}
    {\widehat C}^{(r)}_k=L_k^{(r)\dagger} C^{(r)}_k L_k^{(r)} \ .
    \label{hatCC}
\end{equation}
The circuitlike representation of
$C_{k+1}^{(r)}$ is again given by~\eqref{unitarity1}, but since now $k<M$, all matrices $\bar \Lambda$ are cut away and only remains the network
\begin{equation}
\vcenter{\hbox{\includegraphics[width=0.47\columnwidth]{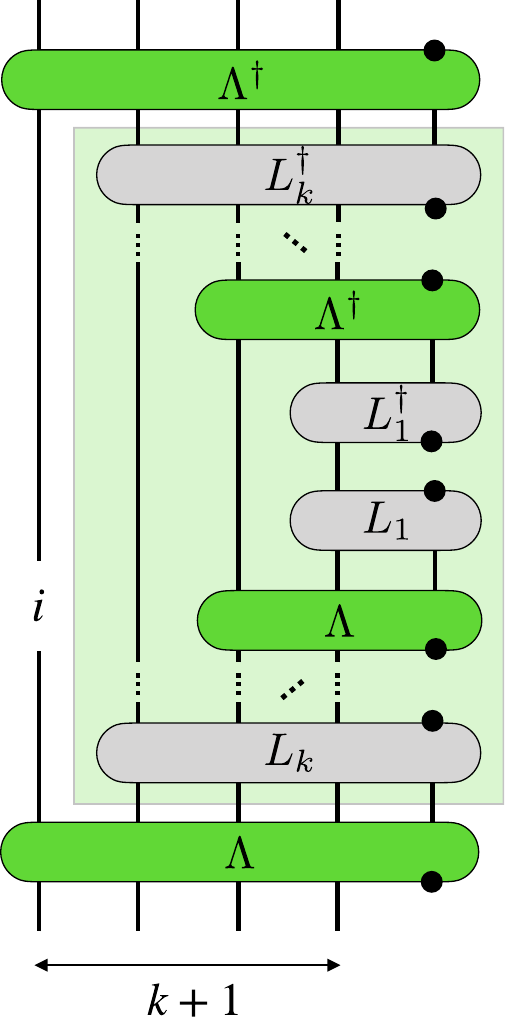}}}  
\label{unitarity2}
\end{equation} 
From~\eqref{hatCC}, the shaded region builds in this case ${\widehat C}_k^{(r-i)}$ and the figure translates into the recursion relations
\begin{equation}
    C^{(r)}_{k+1}=\sum_{i=0,1} \Lambda^{(i,r)\dagger} 
    {\widehat C}^{(r-i)}_{k}\Lambda^{(i,r)} \ .
\end{equation}
When substituted in~\eqref{unit2}, they prove the unitarity of the short gates.

\section{ABA = CBA}
\label{sec:ABACBA}

In this section we will clarify the connection between the approach to derive the Bethe  circuits followed here, based on the CBA, and that in~\cite{Sopena22}, based on the ABA. 

The ABA has a straightforward interpretation as an
MPS, where the associated three-leg tensor is
\begin{equation}
\vcenter{\hbox{\includegraphics[width=0.55\columnwidth]{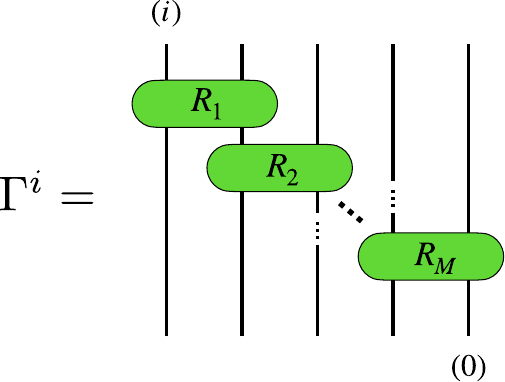}}}  
\label{BA11}
\end{equation} 
This figure makes explicit the decomposition of the tensor ${\mathscr{R}}_T$ in~\eqref{ABAcircuit} in terms of the $R$ matrices of the model.
The XXZ $R$ matrix for the spin-$1/2$ chain is 
\begin{equation}
    R= \rho \begin{pmatrix}
        \,1 \,&0\,&0\,&0\,\\
        0 &y&x&0\\
        0 &x&y&0\\
        0 &0&0&1
    \end{pmatrix}
    \label{R} \ .
\end{equation}
Integrability imposes 
\begin{equation}
    y^2=1+x^2-2 \Delta \, x \ ,
    \label{intc}
\end{equation}
where $\Delta$ is the anisotropy of the Hamiltonian~\eqref{ham}. The global factor $\rho$ is a free parameter, which we will set to $1$. Each $R$ matrix is thus a function of the magnon momentum,  $R_j=R(x_j)$.

Reference~\cite{Sopena22} introduced the transformation of $\Gamma$ into canonical form as the way to derive the quantum gates of the ABC. Let us focus on long gates, since we have seen that short gates involve  additional considerations.
In analogy with~\eqref{ansatz}, long gates were given by 
\begin{equation}
    P_k^{(i,r)}={\mathscr A}_k^{\,(r-i)} \, \Gamma^{(i,r)} {\mathscr B}_{k+1}^{\,(r)} \ ,
    \label{ABAP}
\end{equation}
with ${\mathscr A}_k$ and ${\mathscr B}_k$ inverse of each other~\footnote{This equation corresponds to (17) of~\cite{Sopena22}, with $G_{k-1}$ identified with ${\mathscr A}_k$ and $G_{k}^{-1}$ with ${\mathscr B}_{k+1}$. The matrices $G_k$ were determined there by the recursion relation (16).}. Unlike the comprehensive approach followed here, the
starting point in~\cite{Sopena22} was the recursion relations of the form~\eqref{reclong} under the substitution $\Lambda \to \Gamma$ and $C_k \to {\mathscr C}_k={\mathscr A}_k^\dagger {\mathscr A}_k$.
These recursion relations
were solved iteratively to find the matrix ${\mathscr A}_k$, a task that in general could only be achieved via numerical techniques. 
The ABA and the CBA are known to prepare the same wavefunctions, up to a normalization factor, even before imposing the Bethe equations for the magnon momenta~\cite{Sklyanin82}. 
While it is straightforward to obtain the CBA norms, those of the ABA states pose a notoriously difficult problem~\cite{Korepin82,Korepin1993kvr,Hernandez14}.
With the insights developed in the present paper, the hardness of finding analytic solutions to the ABA recursion relations can be partly explained by this fact. 

Since $\Lambda$ and $\Gamma$ build the same wavefunctions, they must be related by the MPS gauge freedom~\eqref{net3}
\begin{equation}
    \Gamma^{\,i}=  X^{-1} \Lambda^i X \ . 
    \label{AC}
\end{equation}
Given that both tensors are site independent, the transformation $X$ must be site independent. It also has to be consistent with the $U(1)$ symmetry.
The value $\rho=1$ for the global factor of the $R$ matrix leads straightforwardly to $\Gamma^{(0,0)}$=1. The diagrammatic rules constructed in Section~\ref{scattMat} imply $\Lambda^{(0,0)}=1$. This shows that~\eqref{AC} is satisfied in the trivial, zero-charge sector. Choosing a different value for $\rho$ would require introducing a proportionality factor between both sides of~\eqref{AC}. 
We now define
\begin{equation}
    {\mathscr A}_k=A_k X \ , \hspace{8mm} 
    {\mathscr B}_k=X^{-1} B_k \ .
    \label{ABaba}
\end{equation}
It is then clear that
~\eqref{AC} and~\eqref{ABaba} solve the ABA recursion relations, because with these definitions they just reduce to~\eqref{reclong}, and lead to the same quantum gates as~\eqref{ansatz}.

The relation between the tensors $\Gamma$ and $\Lambda$ offers an interesting alternative way to prove the equivalence between the algebraic and coordinate versions of the Bethe ansatz. The existence of a matrix $X$ fulfilling~\eqref{AC} is nontrivial. Let us analyse the two-magnon case as an example.
When $M=2$, we have 
\begin{equation}
\Gamma^0= \begin{pmatrix}
    1 &0&0&0\\
        0 &x_1&y_1 y_2 &0\\
        0 &0&x_2&0\\
        0 &0&0&x_1 x_2
\end{pmatrix}  \ ,
\end{equation}
while $\Lambda^0={\rm diag}(1,x_1,x_2,x_1 x_2)$.
Hence $X$ is a matrix that diagonalizes $\Gamma^0$. This condition only determines $X$ up to left multiplication by a diagonal matrix
\begin{equation}
    X = D \,X^0 \ ,
    \label{DX0}
\end{equation}
with $D={\rm diag}(1,d_1,d_2,d_3)$.
Since $\Gamma^0$ is upper triangular, $X^0$ will also be upper triangular. We fix it by requiring the diagonal entries to be the unity. With this condition, we obtain
\begin{equation}
X^0=\begin{pmatrix}
    1 &\,0&0&0\\
        0 &1&\frac{y_1 y_2}{x_1-x_2} &0\\
        0 &0&1&0\\
        0 &0&0&1
\end{pmatrix}  \ .
\end{equation}
\vspace{1mm}

\noindent
This construction is easily seen to work for arbitrary $M$.

When $i=1$, we have
\begin{equation}
\Lambda^{1}=
\begin{pmatrix}
    0&\;1&\;1&0\\
        0 &0&0&\!-s_{12} x_1\\
        0 &0&0&s_{21} x_2\\
        0 &0&0&0
\end{pmatrix} \ ,
\end{equation}
and
\begin{equation}
\Gamma^{1}=
\begin{pmatrix}
    0&y_1&x_1 y_2&0\\
        0 &0&0&y_2\\
        0 &0&0&x_2 y_1 \\
        0 &0&0&0
\end{pmatrix} \ .
\end{equation}
\vspace{1mm}

\noindent
Equation~\eqref{AC} applied to these matrices overdetermines the diagonal matrix $D$ in~\eqref{DX0}. A solution can only be found if the following relation holds
\begin{equation}
    \frac{s_{21}}{s_{12}}=\frac{x_1 y_1^2-x_1^2(x_1-x_2)}{x_2 y_1^2 +x_1-x_2} \ .
\end{equation}
Exchanging $1 \leftrightarrow 2$ inverts the LHS. It is simple to see that the consistency of the rhs with this property is equivalent to the integrability condition~\eqref{intc}. Then, substituting the integrability condition above, leads to XXZ scattering amplitude~\eqref{sfunction}, up to a symmetric function of the magnon momenta. We will see below that a symmetric $s$ is physically irrelevant and can be thus discarded. Remarkably,~\eqref{AC} encodes the integrability structure of the model, otherwise formulated in terms of the Yang-Baxter equation.

For the sake of completeness, we include the solution of the diagonal matrix 
\begin{equation}
    D={\rm diag}\Big(1,y_1,\frac{ s_{21} y_2}{x_2-x_1}, \frac{ y_1 y_2}{x_2-x_1}\Big) \ ,
\end{equation}
which contains the information about the norms of the ABA states.

\section{\label{sec:free_fermion}Free fermion circuits}

The implementation of our algorithm on a real quantum processor depends on the possibility of efficiently decomposing $P_k$ into elementary quantum gates. 
Here we will only address this very important question when the magnon
$S$ matrix~\eqref{phaseshift} reduces to a sign, whereby the system maps to free fermions through a Jordan-Wigner transformation. 
This corresponds to the symmetric point of the 
scattering amplitude~\eqref{sfunction}, which is achieved for vanishing  $\Delta$ and leads to the XX chain.
At this point, the dependence of the Bethe states on the scattering amplitude factorizes 
\vspace{0mm}
\begin{eqnarray}
  |\Psi_N^{(M)} \rangle &=& \left( \underset{{p>q}}{\prod_{p,q=1}^M} \! \!s_{pq}  \right) \underset{n_l <n_{l\!+\!1}}{\sum_{  n_{l}=1}^N}   \sum_{a_l=1}^M  \epsilon_{a_1 \dots a_M}   \\[2mm]
&&\hspace{-1mm} \times \; x_{a_1}^{n_{1}-1} \! \dots  \, x_{a_M}^{n_{M}-1}  \; | n_1 n_2\cdots n_M \rangle  \ ,\nonumber 
\end{eqnarray}
and thus can be altogether ignored. Without loss of generality, we will set $s_{pq}=1$ in this section. The gates of the ABC at the free-fermion factorize into matchgates, whose analytical expressions we will show now. Strong numerical evidence for this realization was already presented in~\cite{Sopena22}.

\subsection{Matchgate decomposition}

Unitarity strongly constrains the $r=1$ symmetry sector of the ABC gates. Orthonormality between the columns of $P^{(1)}_k$ imposes
\begin{equation}
  P^{(1)}_k =  \begin{pmatrix}
        u_{k1} &  v_{k1}^* u_{k2} & \; v_{k1}^* v_{k2}^* u_{k3} & \dots  \\[2mm]
        v_{k1} &\!\! -u_{k1}^* u_{k2} & - u_{k1}^* v_{k2}^*  u_{k3} & \dots \\[2mm]
        0 & v_{k2} &\!\! -u_{k2}^* u_{k3} & \dots \\[2mm]
        0&0& v_{k3} & \dots \\
        \vdots & \vdots & \vdots
       
    \end{pmatrix} \ .
    \label{nested}
\end{equation}
\vspace{2mm}

\noindent
Following~\eqref{Pni}, the entries in the first row correspond to $P^{(1,1)}_k$, while the rest belong to $P^{(0,1)}_k$. The nested structure of~\eqref{nested} implies that the $r=1$ sector can be represented by a layer of two-qubit unitaries of the form 
\begin{equation}
    F_{kj}=\begin{pmatrix}
    1 & 0 & 0 & 0\\
    0 & u_{kj} & \; v^*_{kj} & 0 \\
    0 & v_{kj} & \! -u^*_{kj} & 0 \\
    0 & 0 & 0 & 1 \\
    \end{pmatrix} \ ,
    \label{matchgate}
\end{equation}
\vspace{2mm}

\noindent
with parameters given by
\begin{equation}
v_{kj}=P^{(0,1)}_{k,jj} \ , \hspace{6mm}
    u_{kj} =\frac{P_{k,1j}^{(1,1)}}{v_{k1}^\ast \ldots v_{kj-1}^*} \ .
    \label{parameters}
\end{equation}

For free fermion models, there is evidence~\cite{Sopena22} that the same layer of $F_{jk}$ unitaries reproduces all symmetry sectors, for which we provide here the analytical expressions. Namely, long and short gates $P_k$ admit respectively the efficient decomposition
\begin{equation}
\vcenter{\hbox{\includegraphics[width=0.82\columnwidth]{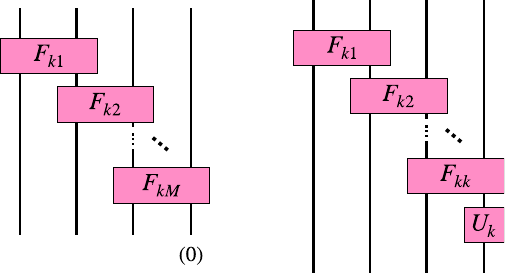}}} 
\label{free1}
\end{equation} 
Although this decomposition appears similar to~\eqref{BA11}, it is considerably more intricate. Contrary to the $R$ matrices, which depend on a single magnon momentum,~\eqref{parameters} implies that $F_{kj}$ depends on $p_1,\ldots,p_j$.
As in the general case, the short gates require adjustments. It is easy to see from~\eqref{nested} that a single qubit unitary acting on the rightmost qubit needs to be added to the $F_{jk}$ layer. The gate $U_k$ is diagonal and acts trivially when the rightmost qubit is in $\ket{0}$. When this qubit is in  $\ket{1}$, it gets multiplied by $u_{kk+1}$ defined as in~\eqref{parameters}, which unitarity ensures to be a phase.

The decomposition~\eqref{free1} means that $r=1$, associated to one-magnon configurations, contains the complete dynamical information of the model. It is a very natural statement in the absence of interaction. Moreover, the matrices $F_{kj}$ are a $U(1)$ preserving restriction of the so-called matchgates~\cite{Josza09,Kraus11}, which are ubiquitous in the circuit realisation of free fermion models.
In spite of that, it turns out that a complete analytical proof of~\eqref{free1} is technically involved.
In~\cite{Sopena22} it was obtained for states with up to three magnons, while successful numerically tests were carried out for a larger number of magnons. A related result holds at the level of overlap matrices.
In the free fermion limit, they
reduce to determinants of inner products of one-magnon states
\begin{equation}
C_{k,ab}^{(r)} = 
\begin{vmatrix}
\, C^{(1)}_{k,n_1 m_1} & \dots & C^{(1)}_{k,n_1 m_r} \\
\, \vdots & & \vdots \\
\, C^{(1)}_{k,n_r m_1} & \dots & C^{(1)}_{k,n_r m_r}
\end{vmatrix} \ ,
\label{Crfree}
\end{equation}
with $a=(n_1,\ldots,n_r)$ and $b=(m_1,\ldots,m_r)$. This equation is just a consequence of Wick's theorem. 
Since overlap matrices are central to the construction of the ABC gates,~\eqref{Crfree} is an important ingredient in the proof of their matchgate decomposition. It is actually the difficulty of operating with the previous expression that prevented us from obtaining the general proof of~\eqref{free1}.

\subsection{Interacting chains}

A first study on the implementation of the ABC for the interacting XXZ model
was performed in~\cite{Sopena22}. The $P_k$ gates preparing states with low numbers of magnons were reproduced via a variational ansatz with several layers of $U(1)$ preserving two-qubit gates. The result of the optimization pointed toward the need of an exponentially growing number of layers with an increasing number of magnons. On the other hand, it is known that the gap for adiabatically building the ground state of the XXZ closes only polynomially~\cite{Yang66}. Since there is a polynomial equivalence between adiabatic and digital computation~\cite{Farhi00}, this suggests that an efficient decomposition of the ABC could exist.

We hope that the structure uncovered in this paper can help to find an improved ansatz.  In this sense, the interest of finding a proof of~\eqref{free1} is not only mathematical. Understanding how the general ansatz~\eqref{ansatz} can be reduced to a single layer of two-qubit unitaries for the XX model, is the natural starting point to address the interacting case.

\section{\label{sec:Conclusions}Conclusions}

In this paper, we have proposed a deterministic quantum circuit designed 
to generate the Bethe wavefunction associated with the XXZ spin chain model. 
Our efforts build upon the approach initiated in reference~\cite{Sopena22}, 
which we now complete
by providing 
comprehensive analytic expressions for the quantum gates, applicable to systems with any number of sites and magnons.
The method employed in 
~\cite{Sopena22} relied on a series of recursive equations 
hinging on the $R$ matrix of the Algebraic Bethe Ansatz which required numerical solutions, particularly for scenarios involving more than two magnons.

In contrast, this paper takes a distinct standpoint that revolves around the discovery of a novel
Matrix Product State structure inherent to the Coordinate Bethe Ansatz. 
This construction offers greater clarity regarding the role 
of the dynamics, as elucidated by the phase and scattering factors of the model.
Furthermore, it possesses an intriguing diagrammatic structure that merits further investigation and exploration. We also highlighted the connection between the method presented here and the approach introduced in~\cite{Sopena22}, based on the Algebraic Bethe Ansatz. As a bonus, we obtain a new understanding of the equivalence between  the coordinate and the algebraic version of the Bethe ansatz.
Looking at the technical aspects, it is worth noting the significant role played by the Cholesky 
decomposition in this construction, which differs from the role undertaken by the QR decomposition in~\cite{Sopena22}.

Our construction does not impose the periodicity condition on the Bethe wavefunction, which is the condition yielding the Bethe equations for the momenta. Consequently, to obtain the eigenfunctions 
of the XXZ Hamiltonian, it becomes necessary to initially solve these equations and then incorporate the solutions into the quantum gates. Alternatively, one could employ the circuit as a tool for a Variational Quantum Eigensolver (VQE) to seek out solutions for the Bethe equations. This presents an interesting avenue for exploration.

A central question regards the decomposition of the ABC gates into one-qubit and two-qubit unitaries. We have strengthened the evidence that an efficient decomposition in terms of matchgates exists for the free fermion chain. Addressing the interacting case is crucial for the implementation of our algorithm on a real quantum computer. At any rate, we should stress that the framework presented here treats free and interacting models on an equal footing, offering analytical expressions for both. It thus appears as an optimal playground to analyse the interplay between integrability, interaction and complexity. 

The complete proof that the gates of the ABC construct actual Bethe wavefunctions is on the other hand missing. 
We have demonstrated that the MPS based on $\Lambda$ yields the CBA. 
We also proved that the canonical form of the MPS brings about the long gates of the ABC.
However, the canonical form does not produce the short gates of the ABC in a straightforward fashion, 
which arise under the elimination of ancillae and render the quantum circuit deterministic.
We have tested that the ABC with short gates prepare the CBA wavefunctions for a few magnons, 
but a complete demonstration would be nonetheless desiderable. Moreover, we have assumed
that the gates of the ABC reduce to one layer of matchgates at the free-fermion point.
We have checked that the matchgate decomposition holds analytically in 
the $r=1$ sector.
Despite strong numerical evidence in favor of the decomposition in every other sector, 
a complete proof still lacks.

Our results  hold the potential for extension in several directions. 
These include exploring open boundary conditions, investigating inhomogeneous systems, 
delving into the nested Bethe ansatz, exploring semi-classical exactly solvable models, and even nonintegrable models.  In this sense, it is interesting to note that the ABC can prepare plane wave superpositions of the form~\eqref{Bwf} for an arbitrary scattering amplitude $s(p,q)$ not related to integrability. 

\section*{Acknowledgements}

We would like to thank 
Francisco Alcaraz, Mari Carmen Banuls, 
Jean Sebastian Caux, 
Ignacio Cirac, Diego Garc{\'i}a-Mart{\'i}n, 
Karen Hallberg, Rafael Hern{\'a}ndez, 
Vladimir Korepin, Barbara Kraus,
Jos{\'e} Ignacio Latorre,
Rafael Nepomechie, Juan Miguel Nieto Garc{\'i}a, Toma{\v{z}} Prosen, 
Kareljan Schoutens
and Peter Zoller for useful
discussions. 

This work has been financially supported by the Spanish Agencia Estatal de Investigacion through the grants “IFT Centro de Excelencia Severo Ochoa
CEX2020-001007-S” and PID2021-127726NB-I00 funded by MCIN/AEI/10.13039/501100011033,
by ERDF,  and the CSIC Research Platform on Quantum Technologies PTI-001. We also acknowledge support 
by the MINECO through the QUANTUM ENIA project call - QUANTUM SPAIN project, and by the EU through the RTRP-NextGenerationEU within the framework of the Digital Spain 2025 Agenda.

R.R. is supported by the UCM, Ministerio de Universidades, and the European Union - NextGenerationEU through contract CT18/22.
A.S. is supported by the Spanish Ministry of Science and
Innovation under the grant SEV-2016-0597-19-4.
M.H.G. was supported by “la Caixa” Foundation (ID
100010434), Grant No. LCF/BQ/DI19/11730056
and by the U.S. DOE, Office of Science, Office of
Advanced Scientific Computing Research, under the
Quantum Computing Application Teams program.

\bibliographystyle{quantum}
\bibliography{main}
\clearpage

\setcounter{equation}{0}\renewcommand\theequation{A\arabic{equation}}

\appendix

\section{
Formulae for the \texorpdfstring{$A$}{A} and \texorpdfstring{$B$}{B} matrices.
}

\label{appAB}

In this appendix, we write down the expressions of $A$ and $B$ in the ansatz~\eqref{ansatz}. We also prove the expression~\eqref{Lmatrix1} of $L$ for short gates. We address the alternatives of long gates~\eqref{net4} and short gates~\eqref{small1} separately.

\subsection{Long gates}

\noindent In Section~\ref{overlapMat}, we defined the $d_{r,M}\times d_{r,M}$ matrix $A_k^{(r)}$ by the Cholesky decomposition~\eqref{Cholesky} of $C_k^{(r)}$, which is the overlap matrix of a set of linearly independent Bethe wavefunctions with $r$ magnons in a spin chain of $k\geq M$ sites.
The fact~$A^{(r)}_k$ is upper triangular and has positive diagonal entries ensures the definition is unique. 

The matrix $B^{(r)}_k$ is the inverse of $A^{(r)}_k$ by definition. This matrix is built upon the Gram-Schmidt process that transforms the nonorthogonal set $|\Psi^{(r)}_{k,a}\rangle$ into a unitary rotation of the orthonormal set $|\Phi^{(r)}_{k,a}\rangle$. Therefore,~$B^{(r)}_k$ admits the closed form
\begin{equation}
\label{Bc}
\begin{split}
B_{k,aa}^{(r)} =\sqrt{\frac{\det_{a-1} C_k^{(r)}}{\det_{a}  C_k^{(r)}}} \ , \quad B_{k,ab}^{(r)} = 0 \! \quad\!  \mathrm{if} \! \quad\! a>b \ , \\ 
B_{k,ab}^{(r)} = -\frac{\det_{b-1} C_{k,a \to b}^{(r)}}{\sqrt{\det_{b-1} C_k^{(r)} \det_{b} C_k^{(r)}} } \quad \mathrm{if} \! \quad\! a<b \ , 
\end{split}
\end{equation}
Let us recall $\det_a M$ denotes the upper left minor of order $a$ of the matrix $M$ and $M_{a\rightarrow b}$ denotes the matrix that results from $M$ under the replacement of the $a$th column by the $b$th column. We shall demonstrate that the closed form of the inverse matrix  to~\eqref{Bc} is
\begin{equation}
\label{Ac}
A_{k,ab}^{(r)} = \frac{\det_{a} C_{k, a \to b}^{(r)}}{\sqrt{\det_{a-1} C_k^{(r)} \det_{a} C_k^{(r)}} } \ .
\end{equation}
The byproduct of~\eqref{Ac} is a determinant formula for the Cholesky decomposition of the overlap matrix (and of any Hermitian, positive definite matrix for that matter). 

Let us alleviate the notation before we address with the proof. Given that we will keep both $r$ and $k$ fixed in the following, 
we introduce the notation
\begin{equation}
    A\equiv A_k^{(r)} \ , \quad B\equiv B_k^{(r)} \ , \quad   C\equiv C_k^{(r)} \ .
\end{equation}
To demonstrate now that $A$ and $B$ above are inverse matrices, we need to check that one of the two alternative products, say $BA$, is proportional to the identity. Furthermore, we just need to focus on the diagonal and upper triangular entries of $BA$, for its lower triangular entries vanish. According~\eqref{Bc} and~\eqref{Ac}, the nontrivial entries of~$BA$ read
\begin{equation}
\delta_{ab}=\frac{\det_{a}\Cc_{a\rightarrow b}}{\det_a\Cc}-\!\!\!\!\sum_{c=a+1}^{b}\!\!\frac{\det_{c-1} \Cc_{a\rightarrow c}\det_c\Cc_{c\rightarrow b}}{\det_{c-1}\Cc\det_c\Cc} \ ,
\label{BA1}
\end{equation}
where $a\leq b$. The identity directly holds if $a=b$. We assume $a<b$ hereafter.

We demonstrate~\eqref{BA1} by proving the lemma
\begin{equation}
\begin{split}
\!\!\!\!\!\!\!\!\frac{\det_{c-1}\!\Cc_{a\rightarrow b}}{\det_{c-1}\!\Cc}\!-\!\frac{\det_{c-1}\!\Cc_{a\rightarrow c}\det_c\!\Cc_{c\rightarrow b}}{\det_{c-1}\!\Cc\det_c\!\Cc}\!=\!\frac{\det_c\!\Cc_{a\rightarrow b}}{\det_c\!\Cc} . \!\!\!\!
\label{lemma}
\end{split}
\end{equation}
If the lemma were to hold, we could start from $c=a+1$ in~\eqref{lemma} and apply it iteratively in~\eqref{BA1} until we prove identity. Our next step exploits the fact $\det_c\Cc_{c\rightarrow b}$ and $\det_c\Cc_{a\rightarrow b}$ share the nonvanishing minor $\det_{c-1}\Cc_{a\rightarrow b}$, whose associated matrix we denote by $M$. The matrix $M$ is nonsingular and invertible because it is built upon the inner product of linearly independent Bethe wavefunctions. We can thus employ formula
\begin{equation}
\label{detid}
\det 
\left( \begin{array}{cc} 
P &  S \\
R & Q \\
\end{array}
\right)   = \det P  \det (Q - R P^{-1} S) \ ,
\end{equation}
to write
\begin{equation}
\label{applid}
\begin{split}
    \frac{{\det}_c \Cc_{c\rightarrow b}}{{\det}_{c-1} \Cc_{a\rightarrow b}}&= -(C_{ca}-w^{\dagger} \,M^{-1} v_{a})\ , \\  
    \frac{{\det}_c \Cc_{a\rightarrow b}}{{\det}_{c-1} \Cc_{a\rightarrow b}}&= C_{cc}-w^{\dagger} \,M^{-1} v_{c}\ , 
\end{split}
\end{equation}
where we have introduced the $c-1$ dimensional column vectors  
\begin{equation}
(v_a)_b=C_{ba} \ , \  w_d=C_{da} \ \, \mathrm{if} \ \, d\neq b \ , \ w_b=C_{ca} \ .
\end{equation}
If we use these expressions and expand $\det_c C$ by minors, we can rephrase~\eqref{lemma} as
\begin{equation}
\begin{split}
\label{int}
    w^{\dagger}\,M^{-1} ({\det}_{c-1}\Cc_{a\rightarrow c} v_a-{\det}_{c-1}\Cc v_{c})\\+\underset{{d\neq a}}{\sum_{{d=1}}^{c-1}}C_{cd}\,{\det}_{c-1}\Cc_{d\rightarrow c}=0\ . 
\end{split}
\end{equation}
The $d$th component of the vector between parentheses consists of two terms of the expansion by minors of $\det_c\Cc^*_{c\rightarrow d}$, which vanishes because of the repetition of rows (recall $d=1,\dots,c-1$) . Therefore,~\eqref{int} equals
\begin{equation}
\begin{split}
    &\underset{{d\neq a}}{\sum_{{d=1}}^{c-1}}{\det}_{c-1}\Cc_{d\rightarrow c}\,(C_{cd}-w^{\dagger}\,M^{-1} v_d)\\
    &=\underset{{d\neq a}}{\sum_{{d=1}}^{c-1}}\frac{{\det}_{c-1}\Cc_{d\rightarrow c}}{\det_c\Cc_{a\rightarrow b}}
    \begin{vmatrix}
    C_{11} & \!\!\dots\!\! & C_{1b} & \!\!\dots\!\! & C_{1d} \\
    \vdots &       & \vdots &       & \vdots \\
    C_{c1} & \!\!\dots\!\! & C_{cb} & \!\!\dots\!\! & C_{cd} \\
    \end{vmatrix}=0 \ ,
\end{split}
\end{equation}
which is an identity due to the repetition of columns in the last determinant. This last step finishes the proof of the lemma~\eqref{lemma}, and, thus, of the identity~\eqref{BA1} and formula~\eqref{Ac}. 

\subsection{Short gates}

\noindent In Section~\ref{sSmallGates}, we defined $A_{k}^{(r)}$ to be $d_{r,k}\times d_{r,k+1}$ matrix that applies the Cholesky decomposition~\eqref{Cholesky2} to~$\widehat{C}_k^{(r)}$\!. The overlap matrix is positive semidefinite rather than positive definitive. The reason is that it is built upon the inner product of linearly dependent Bethe sates. Nonetheless, $A_{k}^{(r)}$ is still unique as it is upper rectangular and has positive diagonal entries in the leftmost square block. This block in fact provides the Cholesky decomposition of $C_k^{(r)}$, the overlap matrix of a complete set of linearly independent Bethe wavefunctions in the $r$th symmetry sector of a spin chain of $k<M$ sites. Since the overlap matrix is now positive definite, we can transfer the conclusions we drew for long gates to this case. The formula for the entries of the leftmost~$d_{r,k}\times d_{r,k}$ block $A_{k}^{(r)}$ is hence~\eqref{Ac}. It also follows that the formula for the~$d_{r,k}\times d_{r,k}$ matrix $B^{(r)}_k$ is~\eqref{Bc} again. (The ranges of the indices of both formulae of course change.) 

To determine $A_k^{(r)}$ completely, we also need to fix the entries beyond the leftmost block. To do so, we look at $|{\Psi}^{(r)}_{k,a}\rangle$, the overcomplete set of $d_{r,k+1}$ Bethe wavefunctions that underpin the positive semidefinite overlap matrix. (We drop the hats from the definitions~\eqref{newstates} and~\eqref{Cholesky2} to avoid cluttering the notation.) We set the first~$d_{r,k}$  Bethe wavefunctions to be a linear basis of~$\mathcal{H}_{r,k}$. This choice implies we can unambiguously express
\begin{equation}
|{\Psi}^{(r)}_{k,b}\rangle=\sum_{a=1}^{d_{r,k}}P_{ab}\,|\Psi^{(r)}_{k,a}\rangle \ , \!\, b=d_{r,k}+1,\dots,d_{r,k+1} \ ,
\label{Psild}
\end{equation}
where $P$ is a rectangular matrix.
The basis of Bethe wavefunctions is amenable to the same reasoning behind formulae for long gates (under the replacement~\mbox{$d_{r,M}\mapsto d_{r,k}$}). Therefore, we can write, up to an unitary rotation, 
\begin{equation}
\begin{split}
|{\Psi}^{(r)}_{k,b}\rangle &= \sum_{a,c=1}^{d_{r,k}} 
\frac{
\begin{vmatrix}
    C_{11} & \!\!\dots\!\! &\!  C_{1a} \! & \!\! P_{cb}C_{1c} \\
    \vdots &       & \vdots & \vdots \\
    C_{a1} & \!\!\dots\!\! & \! C_{aa-1} \! &\!\!  P_{cb}C_{ac} \\
\end{vmatrix}  
}{\sqrt{\det_{a-1} C_k^{(r)} \det_{a} C_k^{(r)}} } |\Phi^{(r)}_{k,c}\rangle  \\
&\equiv\sum_{a=1}^{d_{r,k}}\,A_{k,ab}^{(r)} \,|\Phi^{(r)}_{k,a}\rangle \ , 
\end{split}
\end{equation}
where in the last step we have extended the formula~\eqref{Ac} to $a=1,\dots, d_{r,k}$ and $b=1,\dots,d_{r,k+1}$. The matrix $A_{k}^{(r)}$ thus constructed straightforwardly yields the Cholesky decomposition~\eqref{Cholesky2}. We deduce that~\eqref{Ac} correctly specifies $A_{k}^{(r)}$ for short gates as well.
(The expression~\eqref{Ac} is actually a closed form for the Cholesky decomposition of any positive semidefinite matrix.)

Last, we prove formula~\eqref{Lmatrix1} for $L$. 
Consider the rhs of~\eqref{BA1}. The expression is also valid for~$BA$ in short gates. The difference lies in the ranges of the indices, which now are $a=1,\dots, d_{r,k}$ and~\mbox{$b=1,\dots,d_{r,k+1}$}. If~{$b>d_{r,k}$}, the iteration of lemma~\eqref{lemma} leaves a remnant because it halts at $c=d_{r,k}$. The remnant is just~\eqref{Lmatrix1}, which was to be proven.\\

\section{The tensor network of Bethe wavefunctions}
\label{appBTN}

\setcounter{equation}{0}\renewcommand\theequation{B\arabic{equation}}

In this appendix, we prove that the circuitlike network~\eqref{small9} 
constructs the Bethe wavefunction
\begin{equation}
\begin{split}
\label{Bethewf}
|\Psi^{(r)}_{k,a}\rangle=\!\!\underset{m_b <m_{b\!+\!1}}{\sum_{m_b=1}^{k}}\!\!{\sum_{a_b=1}^r} \epsilon_{m_1 \dots m_r}\Biggl[\,\underset{p>q}{\prod_{p,q=1}^r} \sigma_{a_{p} a_q}\Biggr]\\
\times y_{a_1}^{m_1-1}\ldots y_{a_r}^{m_r-1}\ket{m_1\ldots m_r} \ .
\end{split}
\end{equation}
where $|n_1\ldots n_r\rangle$ is the initial state tensor network,  $y_1,\ldots,y_r$ denote the associated subset of momentum variables of $x_1,\ldots,x_M$ such that $y_a=x_{n_a}$ and $\sigma_{ab}$ denotes the scattering amplitude of $y_a$ and $y_b$. 

Let us begin with the first (lowest) level of~\eqref{small9}. We use $\bar{\Lambda}\ket{0}=\Lambda$,~\eqref{X}, and~\eqref{S} to obtain
\begin{equation}
\begin{split}
\label{lvl1}
\bar{\Lambda}\,|n_{1}\dots n_{r}\rangle\ket{0}
=\Biggl[\,{\prod_{a=1}^r}\,y_a\Biggr]\,\ket{0}|n_{1}\ldots n_{r}\rangle\\+\sum_{a=1}^{r}(-1)^{a+1}\,\Biggl[\,\underset{b\neq a}{\prod_{b=1}^r} \sigma_{ba}y_b\Biggr]\ket{1}|n_{1}\ldots \widehat{n}_a\ldots n_{r}\rangle \ ,
\end{split}
\end{equation}
where $\widehat{n}_a$ denotes $n_a$ is absent in the list.
If we introduce the notation
\begin{equation}
\bar{\Lambda}_j\equiv\mathbb{1}_{2^{j-1}}\otimes\Bar{\Lambda}\otimes\mathbb{1}_{2^{l-j}} \ ,
\end{equation}
where $1\leq l\leq k$, at the second level we have 
\begin{align}
\label{lvl2}
&\bar{\Lambda}_1\,\bar{\Lambda}_2\,|n_{1}\dots n_{r}\rangle\ket{00}=\Biggl[\,{\prod_{b=1}^r}\,y_a^2\Biggr]\,\ket{00}|n_{1}\ldots n_{r}\rangle    \nonumber \\
+&\sum_{a=1}^{r}(-1)^{a+1}\!\,\Biggl[\,\underset{b\neq a}{\prod_{b=1}^r} \,\sigma_{ba}
y_b^2\Biggr]\,(\ket{01}+y_a\ket{10})\,|n_{1}\ldots\widehat{n}_a\ldots n_{r}\rangle \nonumber
\\
-&\underset{a_2\neq a_1}{\sum_{a_2=1}^r}{\sum_{a_1=1}^r}(-1)^{\alpha_1+\alpha_2}\Biggl[\!\!\!\underset{b\neq a_1,a_2}{\prod_{b=1}^r} \!\!\sigma_{ba_2}\sigma_{ba_1}y_b^2\Biggr]\sigma_{a_2 a_1} y_{a_2}|11\rangle \nonumber
\\
\times&\,|n_1\ldots\widehat{n}_{a_1}\ldots\widehat{n}_{a_2}\ldots n_{r}\rangle  \ , 
\end{align}
where $\alpha_1=a_1$ and $\alpha_2$ labels the position of $a_2$ inside the ordered set of $r-1$ elements $1,\ldots,\widehat{a}_1,\ldots, r$. The generalization to the higher levels of~\eqref{lvl1} and~\eqref{lvl2} is apparent. The projection of the last $M$ ancillae on $\bra{0_M}$ at the $k$th level completes the tensor network. 
Therefore, we can write~\eqref{small9} like
\begin{widetext}
\begin{equation}
\label{tn}
\bra{0_M}\bar{\Lambda}_k\ldots\bar{\Lambda}_1|n_{1}\dots n_{r}\rangle\ket{0_k}=\!\!\underset{m_b <m_{b\!+\!1}}{\sum_{m_b=1}^{k}}\underset{a_b\neq a_1,\dots,a_{b-1}}{\sum_{a_b=1}^{r}}\!\!\!\!\!\!(-1)^{\alpha_1\ldots+\alpha_r+r}\Biggl[\,\underset{p>q}{\prod_{p,q=1}^r}\sigma_{a_p a_q}\Biggr]y_{a_1}^{m_1-1}\ldots y_{a_r}^{m_r-1}\ket{m_1 m_2\ldots m_r} \ ,
\end{equation}
\end{widetext}
where $\alpha_b$ labels the position of $m_b$ inside the ordered set of $r-(b-1)$ elements $1,\ldots,\widehat{a}_1,\ldots,\widehat{a}_{b-1},\ldots, r$. The equality between~\eqref{tn} and the Bethe wavefunction~\eqref{Bwf} follows from $\epsilon_{a_1\ldots a_r}=(-1)^{\alpha_1+\ldots+\alpha_r+r}$, which concludes the proof. We close the appendix by stressing that the tensor network in particular prepares~\eqref{Bwf} for $r=M$.

\end{document}